%% file: main.tex
\documentclass[sigconf,nonacm,10pt]{acmart}
\usepackage{graphicx}
\usepackage{caption}
\usepackage{subcaption}
\usepackage{amsmath}
\usepackage[ruled,vlined,linesnumbered]{algorithm2e}
\usepackage[capitalise,noabbrev]{cleveref}
\usepackage{xcolor}
\usepackage{soul}
\usepackage{comment}
\usepackage{nth}
\usepackage{multirow}
\usepackage{makecell}
\usepackage{enumitem}
\usepackage{microtype}
\usepackage{algpseudocode}
\AtBeginDocument{%
  }

\begin{document}

\newbool{showComments}
\booltrue{showComments}

\ifbool{showComments}{%
\newcommand{\cm}[1]{\sethlcolor{yellow}\hl{[Cecilia: #1]}}
\newcommand{\kb}[1]{\sethlcolor{cyan}\hl{[Kayla: #1]}}
\newcommand{\dm}[1]{\sethlcolor{lime}\hl{[Dong: #1]}}
\newcommand{\yl}[1]{\sethlcolor{orange}\hl{[Yang: #1]}}
\newcommand{\ap}[1]{\sethlcolor{red}\hl{[Adam: #1]}}
\newcommand{\qy}[1]{\sethlcolor{green}\hl{[Qiang: #1]}}
\newcommand{\jsw}[1]{\sethlcolor{pink}\hl{[Jake: #1]}}
\newcommand{\mc}[1]{\sethlcolor{red}\hl{[Mathias: #1]}}
}{
\newcommand{\cm}[1]{}
\newcommand{\yl}[1]
\newcommand{\kb}[1]{}
\newcommand{\dm}[1]{}
\newcommand{\ap}[1]{}
\newcommand{\jsw}[1]{}
\newcommand{\mc}[1]{}
}
\newcommand{\systemname}{RespEar\xspace}
\newcommand{\ie}{{\it i.e.,}\xspace}
\newcommand{\eg}{{\it e.g.,}\xspace}
\newcommand{\etc}{\emph{etc.}\xspace}
\renewcommand{\paragraph}[1]{\vspace*{1ex plus 0.25ex minus 0.25ex}\noindent {\bfseries #1}}

\newcommand{\reviewer}[1]{\bf \noindent \large \textcolor{black}{#1}\vspace{0.5pt}\\}
\newcommand{\comm}[1]{\vspace{0pt} \noindent\textcolor{black}{{\bf Comment:} \em #1\\}}
\newcommand{\high}[1]{\vspace{0pt} \noindent\textcolor{red}{#1}}
\newcommand{\resp}[1]{\vspace{0pt}\noindent{\textcolor{black}{\bf  {Response:}}} {\textcolor{black}{#1}\\}}  
\newcommand{\reA}[1]{\noindent {#1}}
\newcommand{\addtext}[1]{\vspace{0pt}\begin{quote} \textcolor{blue}{``#1"}\end{quote}}

\title[RespEar: Earable-Based Robust Respiratory Rate Monitoring]{RespEar: Earable-Based Robust \\ Respiratory Rate Monitoring}

\author{Yang Liu} 
\authornote{Both authors contributed equally to this research.}
\email{yl868@cam.ac.uk}
\affiliation{%
  \institution{University of Cambridge}
  \country{United Kingdom}
}

\author{Kayla-Jade Butkow}
\authornotemark[1]
\email{kjb85@cam.ac.uk}
\affiliation{%
  \institution{University of Cambridge}
  \country{United Kingdom}
}

\author{Jake Stuchbury-Wass} 
\email{js2372@cam.ac.uk}
\affiliation{%
  \institution{University of Cambridge}
  \country{United Kingdom}
}

\author{Adam Pullin}
\email{alp78@cam.ac.uk}
\affiliation{%
  \institution{University of Cambridge}
  \country{United Kingdom}
}

\author{Dong Ma}
\email{dongma@smu.edu.sg}
\affiliation{%
  \institution{Singapore Management University}
  \country{Singapore}
}

\author{Cecilia Mascolo}
\email{cm542@cam.ac.uk}
\affiliation{%
  \institution{University of Cambridge}
  \country{United Kingdom}
}

\renewcommand{\shortauthors}{Yang Liu, et al.}

\input{00-abstract.tex}
\maketitle
\input{01-intro}
\input{02-primer}
\input{03-system}
\input{04-prototyping_data}
\input{05-evaluation}
\input{07-related_work}
\input{08-conclusion}

\bibliographystyle{ACM-Reference-Format}
\bibliography{references}

\end{document}

%% file: 00-abstract.tex
\begin{abstract}

Respiratory rate (RR) monitoring is integral to understanding physical and mental health and tracking fitness. 
Existing studies have demonstrated the feasibility of RR monitoring under specific user conditions (e.g., while remaining still, or while breathing heavily).
Yet, performing accurate, continuous and non-obtrusive RR monitoring across diverse daily routines and activities remains challenging.
In this work, we present \systemname, an earable-based system for robust RR monitoring.
By leveraging the unique properties of in-ear microphones in earbuds, \systemname enables the use of Respiratory Sinus Arrhythmia (RSA) and Locomotor Respiratory Coupling (LRC), physiological couplings between cardiovascular activity, gait and respiration, to indirectly determine RR. 
This effectively addresses the challenges posed by the almost imperceptible breathing signals under daily activities.
We further propose a suite of meticulously crafted signal processing schemes to improve RR estimation accuracy and robustness.
With data collected from 18 subjects over 8 activities, \systemname measures RR with a mean absolute error (MAE) of 1.48 breaths per minutes (BPM) and a mean absolute percent error (MAPE) of 9.12\% in sedentary conditions, and a MAE of 2.28 BPM and a MAPE of 11.04\% in active conditions, respectively, which is unprecedented for a method capable of generalizing across conditions with a single modality. 

\end{abstract}

%% file: 01-intro.tex
\section{Introduction}
\label{s:intro}

RR is a fundamental vital sign that relays pivotal information about health and fitness conditions of the human body. 
Clinically, it is critical for diagnosing and managing various pathologies, acting as an early indicator of health deterioration, such as potential cardiac arrest or respiratory illnesses~\cite{TheImportanceOfRespiratoryRateMonitoringFrom}. In daily life, RR indicates the presence of physical and mental stressors including emotional stress, emotional response, and cognitive load~\cite{TheImportanceOfRespiratoryRateMonitoringFrom}. Additionally, RR is a key indicator of exertion levels during physical activities, offering a valuable insight for managing and optimizing workout routines and detecting exercise-induced fatigue~\cite{FatigueMonitoringThroughWearablesStateoftheArtReview}.
Therefore, continuous RR monitoring across a range of daily settings is essential for gaining valuable insights into health and fitness.

Existing RR monitoring solutions designed to facilitate RR monitoring in daily life primarily rely on three principles: 1) Monitoring breathing-induced body movements. Methods that use IMUs in smartwatches~\cite{BioWatchEstimationOfHeartAndBreathingRates,SleepMonitorMonitoringRespiratoryRateAndBodyPositiona,hao2017mindfulwatch}, smartphones~\cite{rahman2020instantrr,aly2016zephyr,SmartphoneMovementSensorsForTheRemoteMonitoring}, and earbuds~\cite{TowardsRespirationRateMonitoringUsingAnInEar,TowardsMotionAwarePassiveRestingRespiratoryRateMonitoring, RRMonitorResourceAwareEndtoEndSystemForContinuousMonitoring}, require the user to remain still. Pressure sensors in chest straps~\cite{Zephyr} necessitate the use of additional obtrusive wearable devices. Some systems use cameras~\cite{nam2016monitoring} or acoustic sensing~\cite{wang2021smartphone} on smartphones, requiring that the phone be hold in a specific posture and the user remain still. Technologies utilizing wireless signals, such as RF~\cite{ExtractingMultiPersonRespirationFromEntangledRFSignals,SmartHomesThatMonitorBreathingAndHeart} and WiFi~\cite{WiPhoneSmartphonebasedRespirationMonitoringUsingAmbientReflected}are effective under stationary conditions only and are less applicable outdoors.
2) Monitoring breathing-induced airflow through the nose or mouth. These methods involve obtrusive nose-worn sensors~\cite{ContactBasedMethodsForMeasuringRespiratoryRate}, microphones on smartphones positioned near the suprasternal notch and nose~\cite{EstimationOfRespiratoryRatesUsingTheBuiltin}, microphones on earbuds that detect only audible breathing sounds~\cite{InEarAudioWearableMeasurementOfHeartAnda,EstimatingRespiratoryRateFromBreathAudioObtaineda,hu2024breathpro}, or cameras and infrared thermography requiring the user to remain still~\cite{ContactlessVitalSignsMeasurementSystemUsingRGBThermal}.
3) Using behavior or physiological couplings. Previous studies~\cite{RunBuddySmartphoneSystemForRunningRhythmMonitoring,gu2017detecting} exploit the relationship between RR and physical behaviors to indirectly estimate RR, but this method is limited to specific running conditions. Other studies~\cite{natarajan2021measurement,charlton2017breathing,karlen2011respiratory,xu2022toward} have estimated RR from physiological signs, a technique also implemented in commercial smartwatches like Garmin and Apple Watch~\cite{Garmin,AppleWatch}, but only work at rest, \eg ``breathwork'' activities, yoga, or sleeping.

While each existing RR monitoring solution performs well under specific conditions, enabling a unified system through integrating these technologies to operate seamlessly across diverse daily activities is extremely challenging.
It would require not only seamless compatibility across different sensing modalities, but also robust algorithms capable of adapting to diverse environmental factors and user behaviors. 
Moreover, existing solutions often rely on constraints that are not manageable in daily life, such as user stillness or specific device positioning, which complicates their integration into a single versatile system. 
\textit{Therefore, a new approach is needed -- one that transcends the limitations of existing technologies to provide continuous, non-obtrusive RR monitoring that is genuinely effective across a range of daily activities.}

\begin{figure}[t]
    \centering
    \includegraphics[width=1\columnwidth]{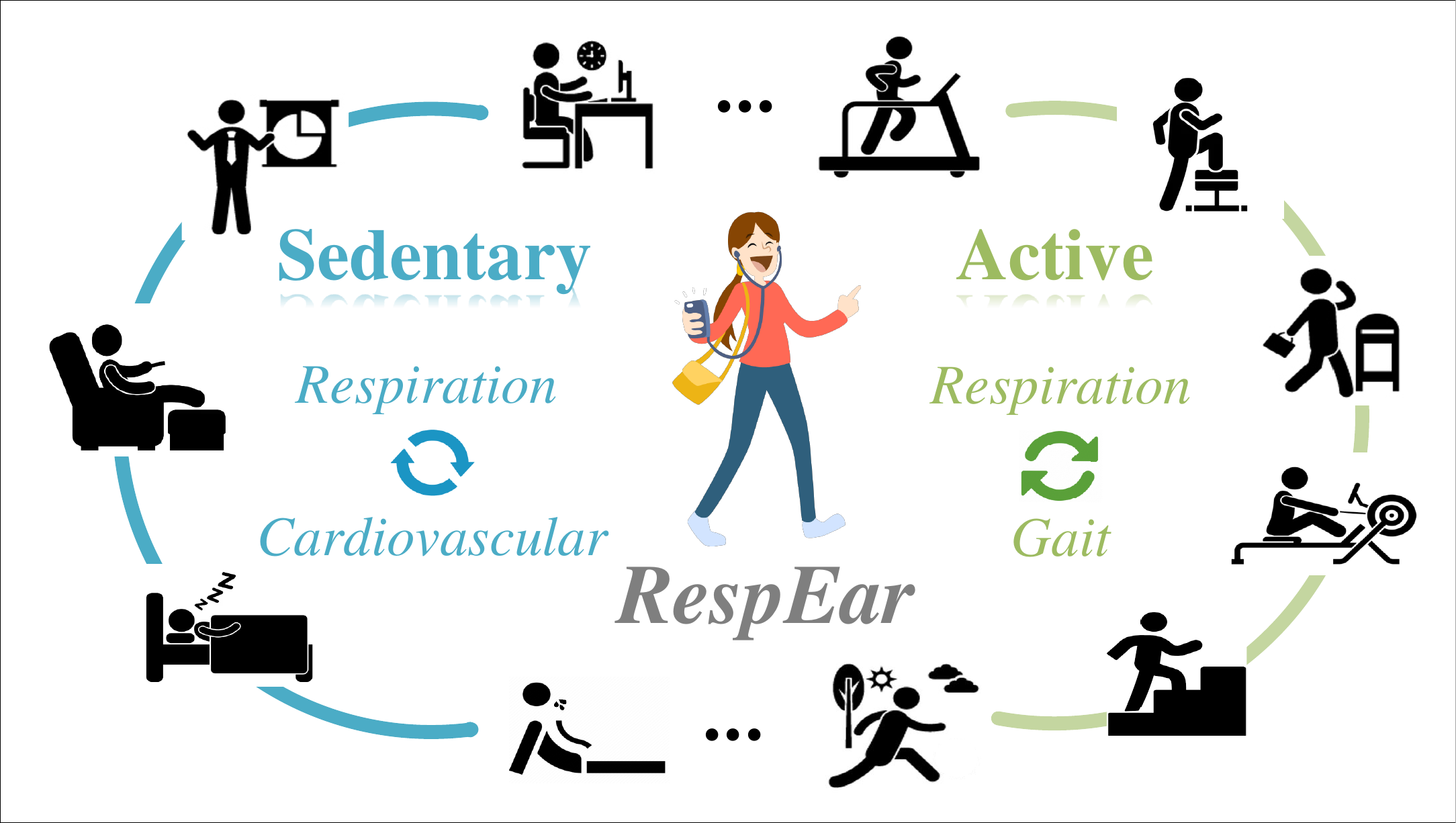}
    \caption{\systemname offers RR monitoring across a variety of daily activities using earbuds.} 
    \label{fig:breath_intro}
\end{figure}

This work presents \systemname, an earable-based system offering robust RR monitoring across both sedentary (\eg sitting, standing, working, cooling down after exercise), and active (\eg walking, running, rowing, and step aerobics) activities as shown in \cref{fig:breath_intro}. 
Considering their widespread popularity and the promising sensing location of the human ear (\ie a stable part of the human body lying near the respiratory organs), we chose earbuds, a mainstream consumer wearable device with daily usage (e.g., entertainment, exercise, and work), as our RR sensing device. 
Moreover, we identified that in-ear microphones are uniquely positioned to measure breathing-related signals (\eg breathing sounds, heartbeats, and footsteps) (\cref{sec:initial_exploration}), thus enabling our solution.

While designing \systemname, we faced these challenges:

\textbf{1) Almost imperceptible breathing sounds.}
The intensity of breathing sounds is minimal when the user is sedentary and overwhelmed by other sounds, like footsteps, when the user is active (\cref{sec:initial_exploration}).
Thus, directly estimating RR using breathing sounds proves unreliable (\cref{s:evaluation}). To address this, we proposed a unified RR monitoring system through identifying in-ear audio for the following purposes:
\begin{itemize}[left=0pt,topsep=0pt]
\item \textit{RSA-based RR monitoring:} When clear heartbeat sounds can be captured using the in-ear microphone (predominantly when the user is sedentary), we can derive heart rate variability (HRV) from in-ear audio. RR is then indirectly estimated using the RSA-based physiological coupling between the cardiovascular activity and respiration, \ie the association between RR and HRV;
\item \textit{LRC-based RR monitoring:} 
However, when clear heartbeat sounds are not available (\eg in the presence of footstep sounds), RSA-based solutions are hindered due to unreliable HRV estimation (validated in \cref{s:evaluation}). Therefore, when rhythmic footsteps are present (i.e., when the user is active), we rely on the in-ear microphone to capture low-frequency footstep sounds, which are used to derive the stride rhythm. Alongside faint high-frequency breathing sounds, RR is estimated by leveraging the LRC-based physical coupling between gait and respiration, specifically the interaction of RR with stride rhythm.
\end{itemize}

\textbf{2) Accurate and reliable estimation.}
Based on the above system, several technical challenges should be addressed to achieve accurate and reliable RR estimation:
\begin{itemize}[left=0pt,topsep=0pt]
\item \textit{RSA-based solutions under varying RRs:}
Practically, RR can change across time, which introduces variability in the association between RR and HRV. This poses challenges in decoupling their relationship.
To overcome this, we formulated an optimization problem aimed at dynamically extracting breathing signals from HRV signals, thereby adapting to these variations to enhance performance.
\item \textit{Respiration-related features extraction for LRC-based RR monitoring:}
Although LRC shows synchronization between stride rhythm and RR, the variable and unknown LRC ratio prevents direct RR estimation from stride frequency alone. It necessitates extracting respiration-related features from in-ear audio, combined with stride rhythm, to accurately estimate RR. However, while the user is under active conditions, faint breathing sounds are heavily interfered with by other sounds, such as footsteps. To precisely extract the breathing-related features and facilitate RR estimation, we propose a scheme to estimate the probability that each audio frame contains breathing. We then applied Singular Spectrum Analysis (SSA)~\cite{elsner1996singular} to the generated probability curve for isolating components related to breathing.
\item \textit{Varying LRC ratios:} 
RR is typically estimated for a window during which LRC ratios may vary, \eg in walking or running by non-regular runners. To address this variability, we propose a method that aggregates breathing-related components from SSA by considering a range of possible LRC ratios, rather than choosing a fixed one.
\end{itemize}
 
We implemented \systemname with an earable prototype and deployed it on an iPhone 12 Pro. \systemname was evaluated across 8 different activities involving 18 subjects, achieving an overall MAE of 1.71BPM and MAPE of 9.68\%, with errors of 1.48BPM (9.12\%) and 2.28BPM (11.04\%) under sedentary and active conditions, respectively. We compared \systemname with recent earable-based solutions with IMUs~\cite{RRMonitorResourceAwareEndtoEndSystemForContinuousMonitoring}, out-ear microphones~\cite{hao2017mindfulwatch,RunBuddySmartphoneSystemForRunningRhythmMonitoring}, and in-ear microphones~\cite{InEarAudioWearableMeasurementOfHeartAnda}, and found that \systemname outperformed in both sedentary and active conditions while additionally being able to cater for both sets of conditions, unlike recent works. 
Additionally, we tested \systemname under a range of realistic conditions such as different noise levels, music playback, in outdoor and indoor environments, in-the-wild, and at different moving speeds. 

In summary, this paper makes the following contributions:
\begin{itemize}[left=0pt,topsep=0pt]
\item It proposes \systemname, the first earable-based system offering continuous and non-obtrusive RR monitoring across diverse daily routines and activities.
\item It leverages solely the in-ear microphone, a sensor naturally present in many earables and puts forward a holistic and optimized solution for RR estimation which leverages intrinsic relationships of our cardiovascular, gait and respiratory systems.
\item It implements a \systemname prototype and describes our extensive dataset and evaluation. Our results demonstrate that \systemname outperforms the state-of-art and is uniquely able to generalize beyond what other systems have been able to do in terms of activity intensity while remaining  robust under different environmental conditions.
\end{itemize}

%% file: 02-primer.tex
\section{Preliminary Investigation}
\label{s:motivation}

\subsection{Initial Exploration: Sensors on Earables}
\label{sec:initial_exploration}

\begin{figure}[t]
    \centering
    \includegraphics[width=1\columnwidth]{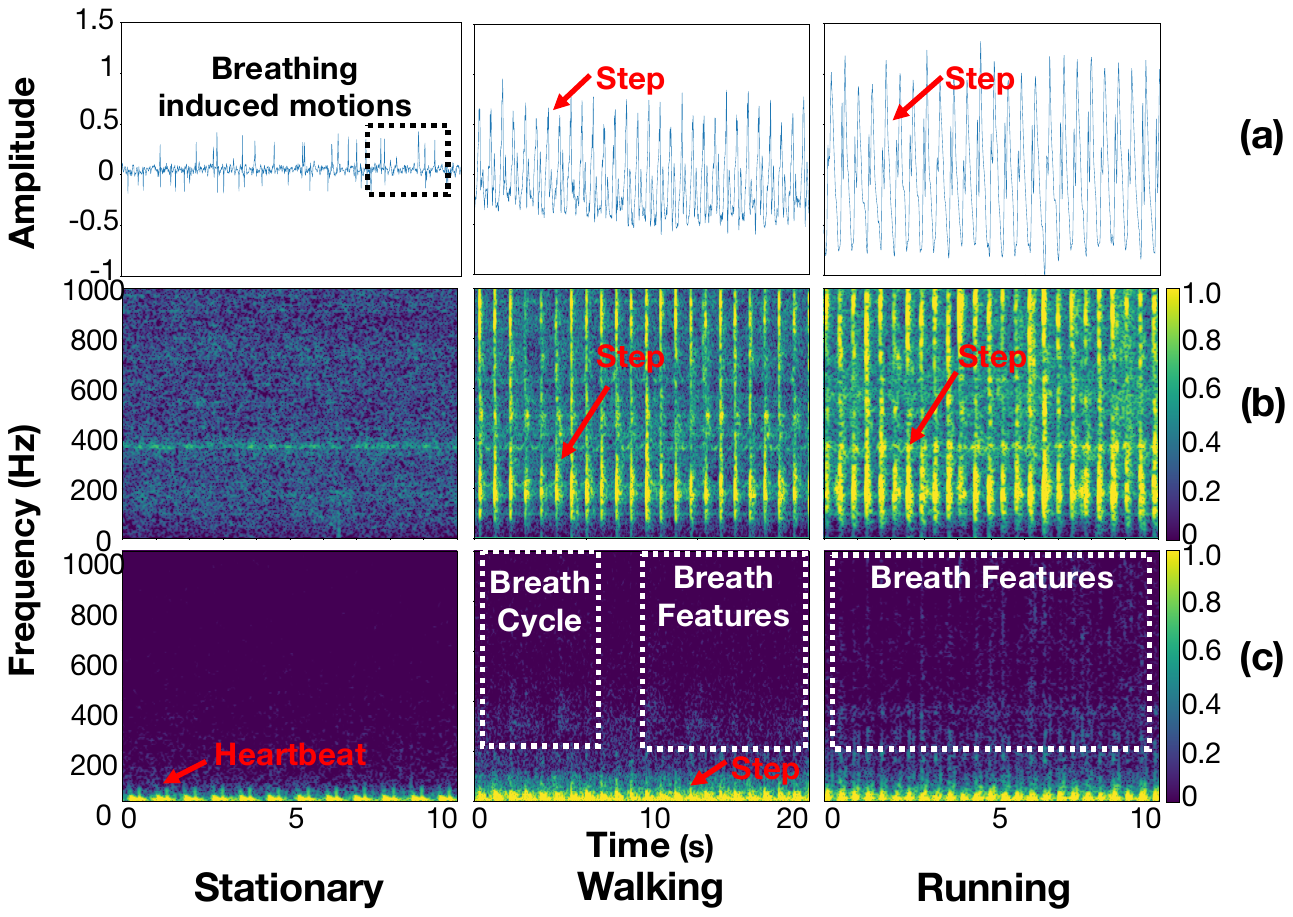}
    \caption{Signals captured using the (a) IMU, (b) out-ear and (c) in-ear microphone under different activities.} 
    \label{fig:preliminary_all_sensors}
\end{figure}

We first explore the feasibility of three common sensors on earables for RR monitoring, namely, \textit{IMU}, used for motion detection and interaction~\cite{EarablesForPersonalScaleBehaviorAnalytics}, \textit{out-ear microphone}, used for speech capture~\cite{EarablesForPersonalScaleBehaviorAnalytics}, and \textit{in-ear microphone}, used for active noise cancellation~\cite{APGAudioplethysmographyForCardiacMonitoringInHearables}. 
We simultaneously collected signals from them when a subject was naturally breathing under three conditions, as shown in \cref{fig:preliminary_all_sensors}: sitting still, walking, and running on a treadmill. The Zephyr BioHarness 3.0 chest strap~\cite{Zephyr} was worn to collect reference signals. 

\textbf{IMU.} We observe from \cref{fig:preliminary_all_sensors}(a) that IMUs can capture breathing-induced motions when the user is stationary, however they are unreliable and have low signal to noise ratio (SNR). Previous studies~\cite{RRMonitorResourceAwareEndtoEndSystemForContinuousMonitoring, TowardsRespirationRateMonitoringUsingAnInEar} apply a band-pass filter (BPF) on the IMU signals while stationary, achieving acceptable performance as validated in \cref{s:evaluation}. However, in continuous practical conditions, the subtle breathing signals are easily overwhelmed by head motions, resulting in poor data retention~\cite{TowardsMotionAwarePassiveRestingRespiratoryRateMonitoring}. When walking or running, the IMU clearly detects steps, but breathing signals are indiscernible, since walking and running generate strong motions that completely overshadow the breathing signal~\cite{castillo2018shock}.

\textbf{Out-ear Microphone.} As spectrograms shown in \cref{fig:preliminary_all_sensors}(b), the out-ear microphone only captures environmental noises under stationary, failing to detect breathing sounds due to large attenuation of faint breathing in air. During walking or running, footstep sounds can be detected (depending on shoe type and ground material), while breathing sounds are not discernible due to attenuation of breathing in air, the loud footstep sounds, and the ambient noise of the treadmill.

\textbf{In-ear Microphone.} As spectrograms shown in \cref{fig:preliminary_all_sensors}(c), under stationary conditions, the faint breathing sounds are undetectable by in-ear microphones. However, sounds of heartbeats are clearly captured due to the occlusion effect~\cite{butkow2023heart}.
When walking or running, we observe that 1) footsteps can be clearly detected because of the occlusion effect 
\cite{OESenseEmployingOcclusionEffectForInearHuman};
2) due to the variations in breathing intensity, only certain breathing sounds are discernible, while the full breathing cycle cannot be distinguished; 3) the in-ear microphone is resilient to ambient noise as it resides inside the ear canal.

\textbf{Summary.} The IMU works in stationary conditions but is sensitive to motion. While the out-ear microphone is vulnerable to ambient noise under all conditions and merely get noise while stationary. The capability of the in-ear microphone to capture heartbeats while stationary, footsteps and partly discernible breathing during walking/running shows the potential to offer solutions for RR monitoring that work effectively in various conditions. Therefore, we select the in-ear microphone as the sensing modality in \systemname.

\subsection{Design Primer}
\label{sec:design_primer}

In-ear microphones can capture versatile audio data in different conditions, yet the method to accurately correlate these signals with RR estimation remains unclear. Inspired by the physiological couplings between cardiovascular activity and respiration~\cite{natarajan2021measurement}, \eg RSA, we are initially exploring methods grounded on this physiological principle.

\subsubsection{RSA-based RR Estimation}
\label{sec:RSA}

RSA is the natural variations in HR that occur due to synchronization with the respiratory cycle~\cite{natarajan2021measurement}. Due to RSA, HR increases during inhalation and decreases during exhalation, resulting in a breathing related modulation of the HRV (\ie the variation in time interval between successive heart beats)~\cite{yasuma2004respiratory}. Rhythms in the low frequency (LF) range of the HRV, spanning from near 0.04 to 0.15 Hz, serve as indicators of sympathetic modulation~\cite{aysin2006effect}. Those within the high frequency (HF) range (near 0.15 to 0.4 Hz) encapsulate rhythms governed by parasympathetic activity, which is closely related to respiration~\cite{aysin2006effect}.

The clearly captured heartbeat sounds from in-ear microphones under sedentary conditions offer the possibility of monitoring RR using RSA.
RSA exists under intense full body motions~\cite{InfluencesOfBreathingPatternsOnRespiratorySinus} (\ie in active conditions); however, accurately extracting heartbeat locations from in-ear audio for HRV estimation in such conditions remains an unsolved and challenging issue~\cite{butkow2023heart,BUTKOW2024101913,EarmonitorInearMotionresilientAcousticSensingUsingCommodity}, as validated in \cref{s:evaluation}.
Thus, it is necessary to identify an alternative method for RR monitoring in active conditions.

\subsubsection{LRC-based RR Estimation}
\label{sec:LRC}
We observe that audio from in-ear microphones under active conditions, such as walking, running, or other activities with rhythmic footsteps, is dominated by footstep sounds and breathing sounds are not always discernible (\cref{sec:initial_exploration}).
Thus, we explore the physiological couplings between gait and respiration here.

LRC is a universal phenomenon in activities that produce and utilize energy rhythmically~\cite{hoffmann2012sound}, such as walking, running, swimming, and rowing~\cite{TheInteractionsBetweenLocomotionAndRespiration,DoesColdWaterEnduranceSwimmingAffectPulmonaryFunction,LocomotorrespiratoryCouplingDevelopsInNoviceFemaleRowers,LocomotorrespiratoryCouplingPatternsAndOxygenConsumptionDuring}. It reveals the interconnected dynamics between RR and stride rhythm~\cite{ImpactLoadingAndLocomotorRespiratoryCoordinationSignificantlyInfluence}, indicating the synchronization between an individual's stride rhythm and their RR. This implies that there will normally be a certain number of steps for each breath (\ie inhalation or exhalation).
In human locomotion, a number of LRC ratios are observed, \eg 4:3, 3:2, 2:1, where an LRC of 2:1 means two steps are taken for one breath.
Thus, the in-ear audio, containing clear footsteps and partly discernible breathing sounds, offers the possibility for  estimating RR based on LRC.

%% file: 03-system.tex
\section{System Design}
\label{s:design}

\begin{figure}[th]
    \centering
    \includegraphics[width=1\columnwidth]{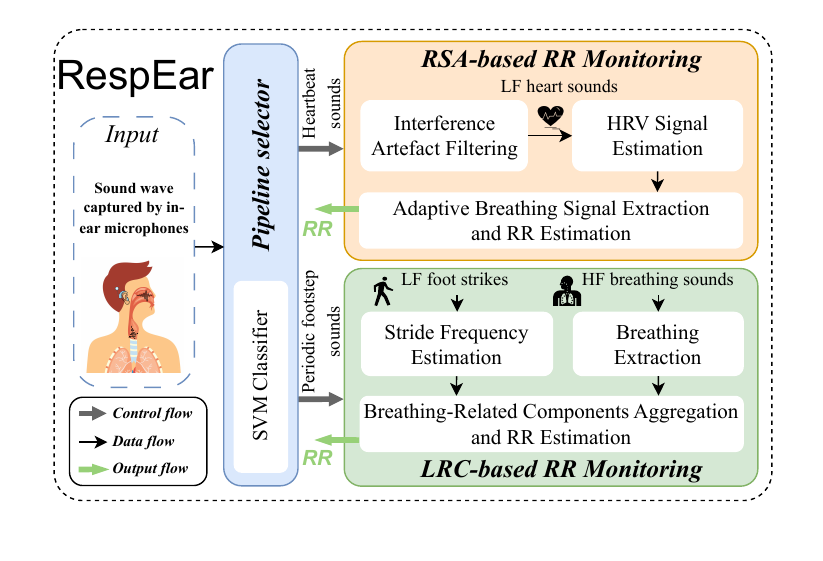}
    \caption{Illustration of the \systemname architecture.}
    \label{fig:system_architecture}
\end{figure}
 
\cref{fig:system_architecture} shows the system architecture. 
\systemname uses 60s estimation windows with 30s overlap to produce a RR estimate per window.
\systemname uses two paths for RR estimation depending on the presence of rhythmic footsteps or clear heartbeat sounds in the audio. If clear heartbeat sounds are present (\ie sedentary conditions), RR will be estimated through the \textbf{RSA-based RR monitoring} pipeline. If rhythmic footsteps are present (i.e., active conditions), RR will be estimated using the \textbf{LRC-based RR monitoring} pipeline.

\subsection{RSA-based RR Monitoring}
\label{s:design:stationary}

\subsubsection{Design Principle}
\label{s:design:stationary:prin}

The high-level process of RSA-based RR monitoring can be summarized in three steps: 
\textbf{1) HRV signal estimation}: Heartbeat locations are detected, and the HRV signal is computed as the time difference between successive heartbeats. \textbf{2) Breathing signal extraction}: A bandpass filter (BPF) is applied to the HRV signal to capture respiration-related rhythms, extracting the high-frequency (HF) range as the extracted breathing signal. \textbf{3) RR estimation}: The final RR is determined by either applying peak detection to the extracted breathing signal and counting the peaks or by using Fast Fourier Transform (FFT) to identify the frequency component with the largest peak.
 
\textbf{Observation.} 
Prior RSA-based methods following the above process, primarily applied to photoplethysmogram (PPG) or electrocardiogram (ECG) signals~\cite{natarajan2021measurement,schafer2008estimation,karlen2011respiratory,xu2022toward}, typically use default and fixed cutoff frequencies for the BPF.
However, we observe that using a fixed frequency range leads to sub-optimal or inaccurate localization of respiration-related rhythms (\ie HF range) in the HRV signal and therefore poor RR estimation. Since the HF range should be centered around the true RR, when the true RR changes over time, the HF range should change accordingly \cite{aysin2006effect}.

\begin{figure}[t]
    \centering
    \includegraphics[width=1\linewidth]{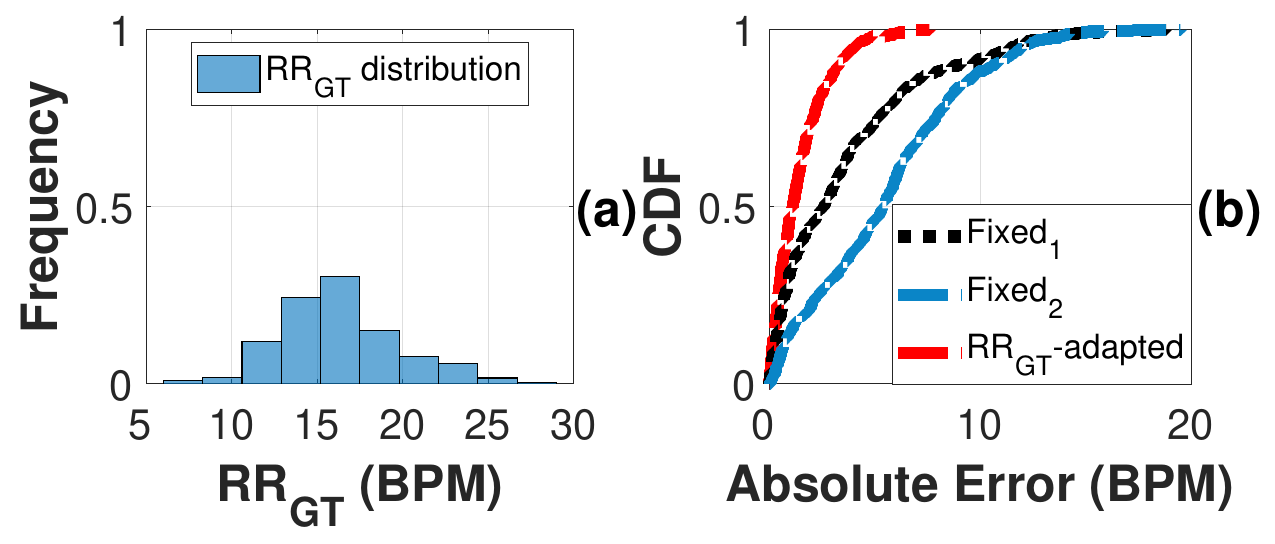}
    \caption{(a) Distribution of $RR_{GT}$ while sedentary. (b) Comparison of RR estimation performance using different BPFs.}
    \label{fig:design:cutoff_showcase1}
\end{figure}

We conducted a study to validate our observation using all of our collected in-ear audio data with clear heartbeat sounds (\ie while sedentary), where the distribution of the Ground Truth (GT) RR (denoted as $RR_{GT}$) is shown in \cref{fig:design:cutoff_showcase1}(a). Specifically, we follow the process of RSA-based RR monitoring, whereby we determine RR by counting the peaks of the extracted breathing signal. The HRV signal is calculated as detailed in \cref{sec:hrv}. 
We compare the RR estimation performance among: (i) the BPF using the default and fixed frequency range in~\cite{morales2020model,xu2022toward}, \ie $Fixed_1 = [0.15,0.35]Hz$, (ii) the predetermined BPF covering the full frequency range of the $RR_{GT}$, \ie $Fixed_2 = [0.1,0.5]Hz$, (iii) the BPF using an adaptive frequency range, \ie $RR_{GT}$-adapted, calculated as $[0.65*RR_{GT}, 1.35*RR_{GT}]$~\cite{aysin2006effect}. The resultant performance is shown in \cref{fig:design:cutoff_showcase1}(b). It is evident that there is a large performance gain by using the $RR_{GT}$-adapted frequency range (MAE = 1.45BPM) compared to the $Fixed_1$ used in previous studies (MAE = 3.54BPM), and the $Fixed_2$ covering all frequencies of $RR_{GT}$ (5.45BPM). This is because the fixed BPF is effective only if $RR_{GT}$ falls within the BPF range, but even then, a non-centered HF range can degrade the performance.

\paragraph{Our design.} The use of adaptive HF range localization on the HRV signal demonstrates the potential for significant performance improvement. We propose a novel approach whereby we formulate and solve an optimization problem to dynamically localize the HF range. To the best of our knowledge, \systemname is the first work to achieve dynamic HF range localization for RSA-based RR estimation. We believe our methodology could also benefit other RSA-based solutions using various sensing modalities, \eg ECG and PPG. We also propose a series of techniques to enable the full pipeline for RSA-based RR monitoring using in-ear audio.

\subsubsection{HRV Signal Estimation.}
\label{sec:hrv}

\begin{figure}[t]
    \centering
    \includegraphics[width=1\linewidth]{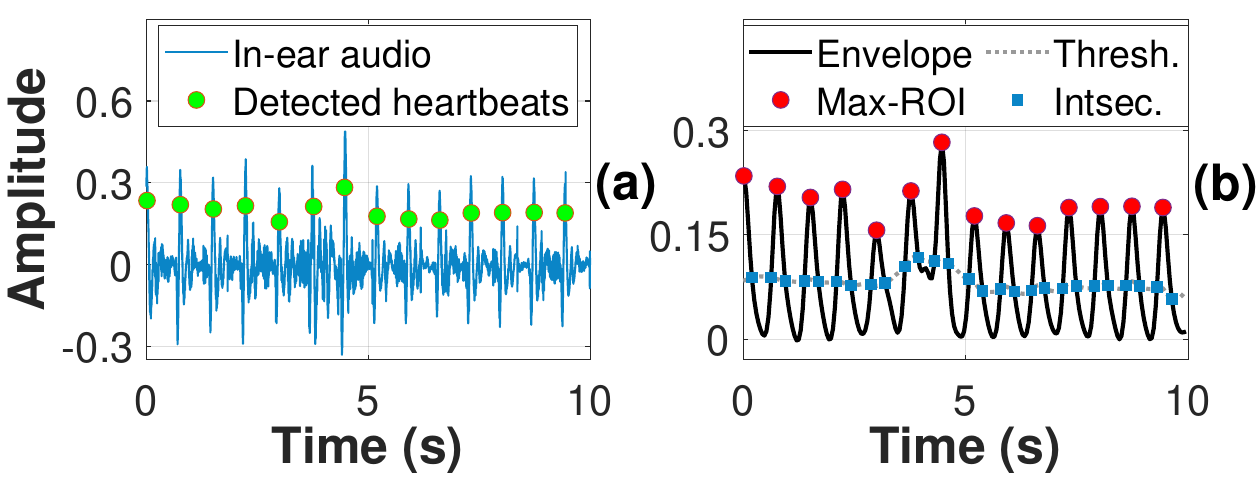}
    \caption{(a) In-ear audio and heartbeats. (b) Peak detection.}
    \label{fig:design:heart_sound}
\end{figure}

\textbf{Heartbeats detection.} A low-pass filter with 30Hz cutoff is used to remove high frequency noise from the in-ear audio, as heartbeats are low-frequency sounds~\cite{butkow2023heart} (\cref{fig:preliminary_all_sensors}). To obtain the HRV signal (denoted as $S_{HRV}$), the heartbeats need to be identified by detecting the peaks in the filtered audio (\cref{fig:design:heart_sound}(a)). To accurately detect peaks while accommodating variations in amplitude and morphology changes, \systemname uses peak detection with an adaptive peak detection threshold. As shown in \cref{fig:design:heart_sound}(b), we first compute the smoothed Hilbert envelope of the filtered audio.
Then, we compute a moving average for each point of the envelope which acts as an adaptive threshold (Thresh.). Regions of interest (ROI) are picked between each pair of intersection points of the envelope and the threshold. Heartbeat peaks are marked at the maximum of each ROI (Max-ROI) between two intersection points (Intsec.) if their amplitude is larger than that of the two enclosing intersection points. $S_{HRV}$ is calculated by the time difference between successively detected peaks (\cref{fig:design:searching}(a)).

\textbf{Automatic channel selection.} Leveraging the unique ability of earables to give two in-ear audio channels (\ie left and right ear), \systemname automatically selects the channel with lower standard deviation (STD) in the estimated HRV signal from each ear. This is because heartbeats are regular signals, so the lower STD implies a less noisy and more robust signal due to more regular heartbeat peaks.

\subsubsection{Adaptive Breathing Signal Extraction.}

\textbf{Design insight.} To enable adaptive HF range localization on $S_{HRV}$, we formulate an optimization problem based on the following observations: First, given a list of potential RR candidates, the best candidate is the one closest to $RR_{GT}$. Second, the optimal RR candidate is the one that, when used to set the frequency range of the BPF, yields the best estimate of $RR_{GT}$. In other words, the best RR candidate minimizes the difference between itself and the RR estimated using the BPF with a frequency range centered around the candidate. Therefore, we select the RR candidate with the minimum difference from the list of possible candidates. 

\begin{figure}[t]
    \centering
    \includegraphics[width=1\linewidth]{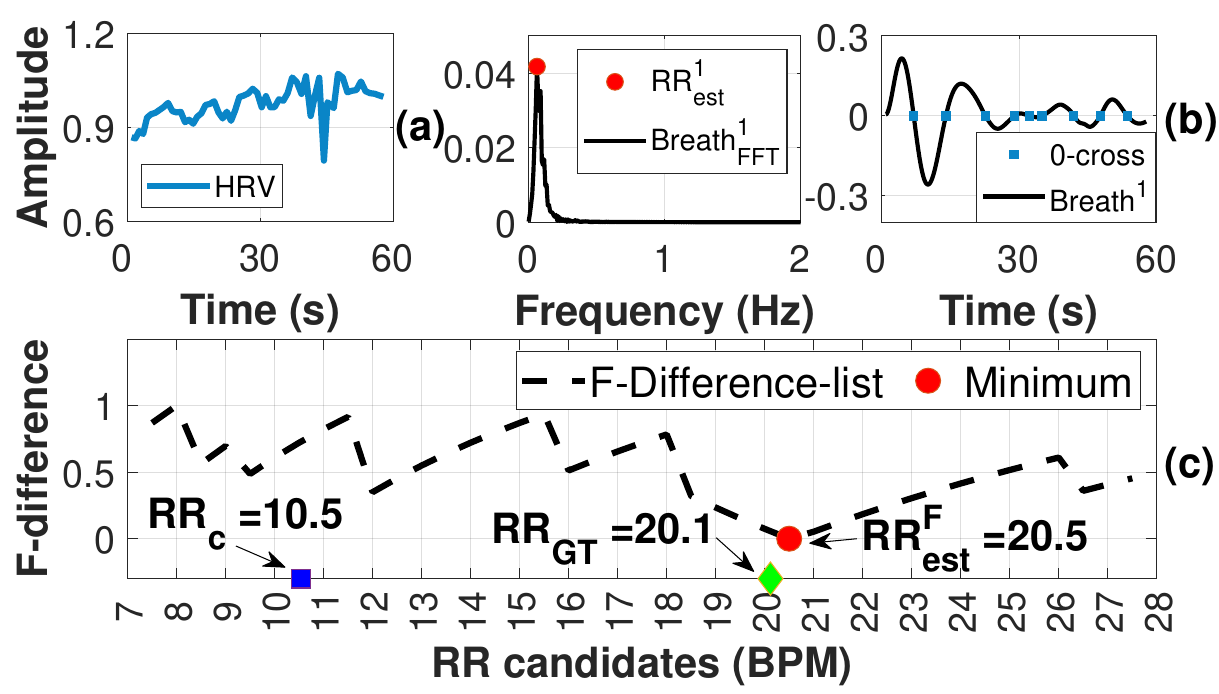} 
    \caption{(a) Extracted HRV signal. (b) FFT determination and zero crossing counting. (c) Best RR candidate searching.}
    \label{fig:design:searching}
\end{figure}

\begin{figure}[th]
    \centering   
    \includegraphics[width=1\linewidth]{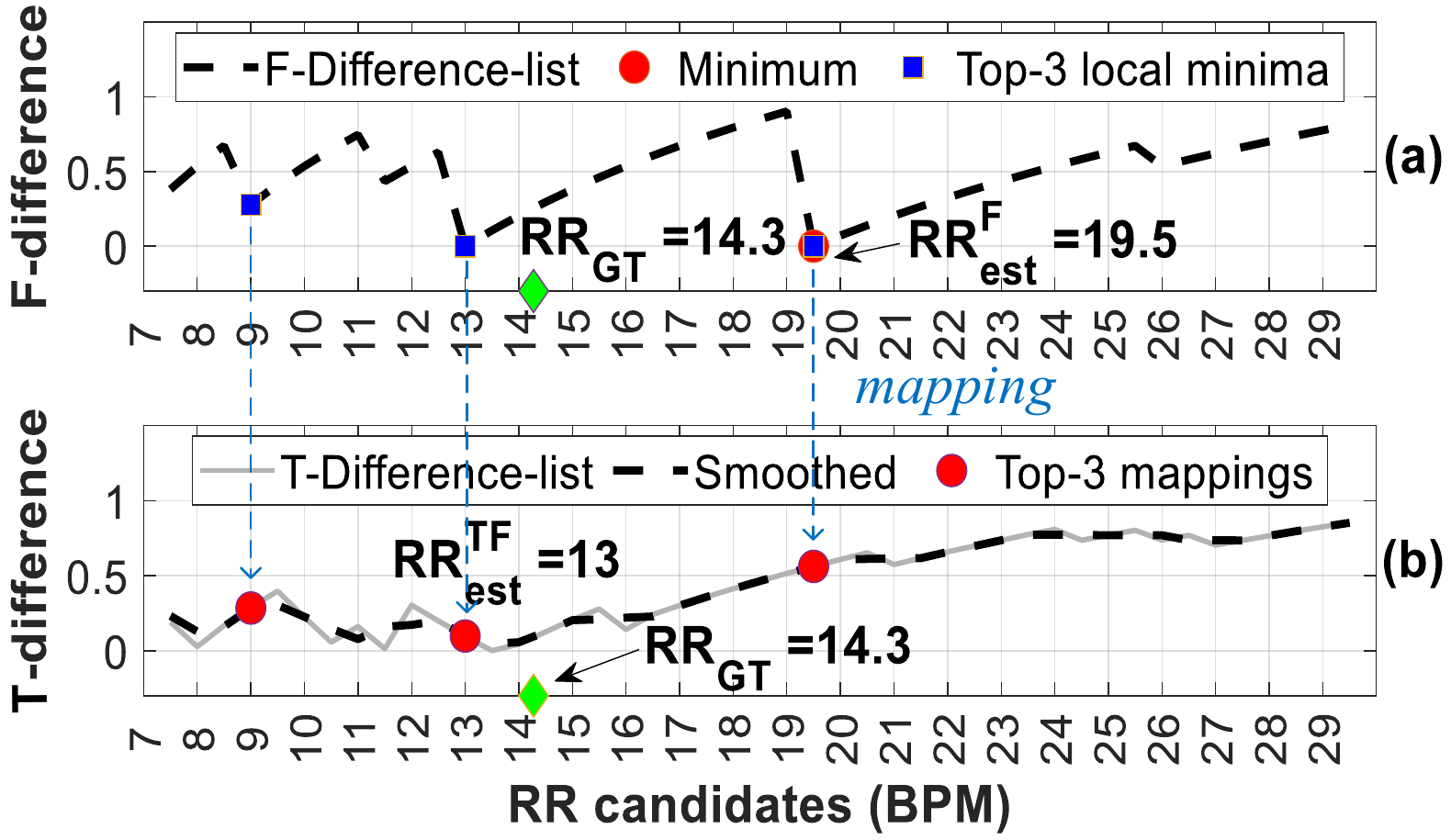} 
    \caption{(a) Best RR candidate searching on the frequency domain. (b) Calibration from the time domain.}
    \label{fig:design:searching_2}
\end{figure}

\textbf{Our algorithm.} We elaborate our algorithm as follows:
\begin{itemize}[left=0pt,topsep=0pt]
\item RR candidate sampling: Sampling RR candidates to generate a list of potential candidates, as detailed in \cref{alg:rr-selection}.
\item Best RR candidate search: Finding the best RR candidate ($RR^F_{est}$) from the generated F-difference-list (\cref{alg:rr-selection}).
\item Calibration from time domain: As shown in \cref{fig:design:searching_2}(a), sometimes we find $RR^F_{est}$, yet is not the optimal RR candidate, likely due to a lower quality $S_{HRV}$. To enhance robustness, we repeat \textit{Step 2) in \cref{alg:rr-selection}} using the RR of each extracted breathing signal estimated in the time domain by counting the zero-crossing points, as shown in \cref{fig:design:searching}(b). This generates a time difference list (T-Difference-list).
We observe that the optimal RR candidate typically appears among the three candidates that have the smallest local minima in the F-Difference-list (\cref{fig:design:searching_2}(a)). Furthermore, the T-Difference-list suggests the trend leading towards the optimal candidate. Therefore, we first smooth the T-Difference-list to highlight its underlying trends, and then sum the T-difference value and the F-difference value of these three candidates respectively. We select the candidate with the smallest sum as the estimated RR as depicted in \cref{fig:design:searching_2}(b).
\end{itemize}

\begin{algorithm}[t]
\caption{Adaptive Breathing Signal Extraction}
\label{alg:rr-selection}
\KwIn{$S_{HRV}$: Heart rate variability signal (\cref{fig:design:searching}(a))}
\KwOut{$RR^{TF}_{est}$: Estimated respiratory rate}
\BlankLine

\begin{algorithmic}[1]
    \item[1.] \textbf{Step 1: RR candidates sampling: $RR_{list}$ (\cref{fig:design:searching}(c))}
    \begin{algorithmic}[1]
    \item[1)] \textbf{Calculating the centre RR of $RR_{list}$: $RR_c$}
            \item[-] Filter $S_{HRV}$ using BPF with cutoffs $[0.15, 0.35]$ Hz
            \item[-] Perform FFT on filtered signal to find the frequency component with the highest amplitude
            \item[-] Convert this frequency to BPM to obtain $RR_c$
    \end{algorithmic}
    
    \begin{algorithmic}[1]
    \item[2)] \textbf{Generating $RR_{list}$}
            \item[-] Identify the smallest and largest human RR~\cite{BRRange}: $7.5BPM$ and $42.5BPM$ respectively
            \item[-] Sample $RR_{list}$ in increments and decrements of 0.5 BPM, where the minimum and maximum RR candidate satisfies:
            \[
            \min(RR_{list}) = \max(7.5, RR_c - w/2)
            \]
            \[
            \max(RR_{list}) = \min(42.5, RR_c + w/2)
            \]
            where $w$ is the predefined length of $RR_{list}$
    \end{algorithmic}
\end{algorithmic}

\begin{algorithmic}[1]
    \item[2.] \textbf{Step 2: Best RR candidate search: $RR^F_{est}$ (\cref{fig:design:searching}(c))}
    \begin{algorithmic}[1]
    \item[1)] \textbf{for} each $RR_{list}^i$ in $RR_{list}$ \textbf{do}
                \item[-] Set BPF with cutoffs $[l^i, h^i]$ Hz, where:
                \[
                l^i = 0.65 \cdot RR_{list}^i / 60; h^i = 1.35 \cdot RR_{list}^i / 60
                \]
                
                \item[-] Filter $S_{HRV}$ with $[l^i, h^i]$ to obtain its extracted breathing signal: $Breath^i$

                \item[-] Estimate the RR of $Breath^{i}$ using the FFT to find the frequency component with the largest peak: $RR_{est}^i$ (\cref{fig:design:searching}(b))

                \item[-] Calculate the difference: $|RR_{est}^i - RR_{list}^i|$
    \end{algorithmic}

    \begin{algorithmic}[1]
    \item[2)] Get the frequency difference list of $RR_{list}$ from 1)
    \end{algorithmic}

    \begin{algorithmic}[1]
    \item[3)] Min-Max normalize the frequency difference list: F-Difference-list (\cref{fig:design:searching}(c))
    \end{algorithmic}

    \begin{algorithmic}[1]
    \item[4)] The candidate with the minimum difference is selected as the best RR candidate
    \end{algorithmic}
    
\end{algorithmic}

\begin{algorithmic}[1]
    \item[3.] \textbf{Step 3: Calibration from time domain: $RR^{TF}_{est}$}
\end{algorithmic}

\end{algorithm}

\subsubsection{Interference Artefact Filtering.}
\label{s:design:stationary:ma}

While the body is undergoing non-full body motions (\ie the user is sedentary), in-ear audio is prone to  interference artefacts. Examples of these include head motions (drinking, speaking\footnote{Breathing adapts to speaking (inhalation at syntactic pauses and exhalation during speech \cite{fuchs2021respiratory}), eliminating the need for RR estimation while speaking.}, motions of the arms, movement of the trunk, respiratory-related sounds (swallowing, coughing, \etc)). 
To ensure accurate sedentary RR estimation even in the presence of interference artefacts, we present an interference filtering approach.

\textbf{Interference artefact detection.} Without interference artefacts, in-ear audio maintains a consistent waveform with clear heart sounds and thus stable statistics over time. 
Conversely, artefacts, such as one-time head motion, cause significant statistical variations. Hence, we propose a statistics-based approach to detect the presence of artefacts. For each 60s window, we segment the audio signal into 3s segments. We compute STD of each segment (to measure signal dispersion) and if it is larger than an empirical threshold, this segment is marked as an interfered segment.

\textbf{Adaptive filter.} If an interfered segment is detected, we use an adaptive filter implemented using the recursive least squares (RLS) algorithm \cite{cioffi1984fast} to remove its interference artefact. The RLS algorithm recursively finds filter coefficients that minimize a least squares cost function with a reference signal. For the interfered segment, we select the nearest segment without interference as the reference signal. After filtering, the interference artefact of the segment are mostly removed for reliable RR estimation (as validated in \cref{s:evaluation}).

\subsection{LRC-based RR Monitoring}
\label{s:design:moving}

\begin{figure}[t]
    \centering
    \includegraphics[width=1\linewidth]{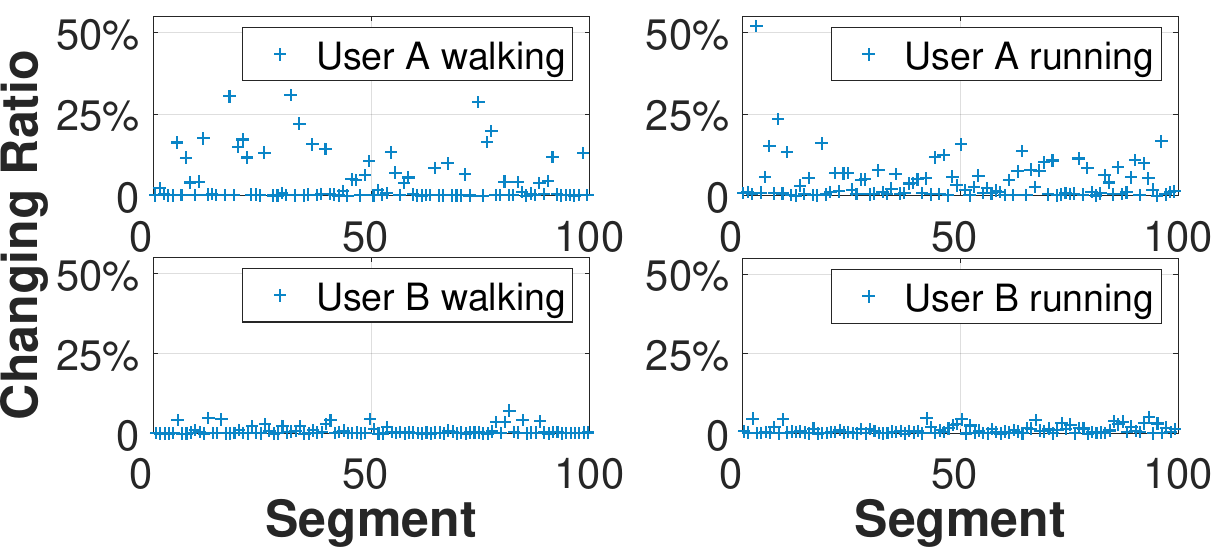}  
    \caption{Changing ratio of GT LRC from two users.}
    \label{fig:design:LRC_showcase1}
\end{figure}

\textbf{Design Challenges.}
Although LRC indicates synchronization between stride rhythm and RR, the LRC ratio between stride rhythm and RR is variable and unknown. Consequently, RR cannot be directly estimated from stride frequency alone. Inspired by previous studies~\cite{RunBuddySmartphoneSystemForRunningRhythmMonitoring, gu2017detecting} which linked stride frequency with breathing signals to estimate the LRC ratio (but only under  running scenarios with a fixed LRC ratio per window), we propose our pipeline. This pipeline addresses two unique challenges identified in \systemname working scenarios:
\begin{itemize}[left=0pt,topsep=0pt]
\item Respiration sounds are strongly interfered by other sounds in in-ear audio, especially footstep sounds which are strong and amplified due to the occlusion effect~\cite{OESenseEmployingOcclusionEffectForInearHuman} (\cref{fig:preliminary_all_sensors}).

\item The LRC ratio varies within an estimation window, \eg for walking and non-regular runners, as demonstrated in \cref{fig:design:LRC_showcase1}.
We analyze in-ear audio from two participants—User A (a non-regular runner) and User B (a regular runner)—by segmenting their audio into 10-second intervals and calculating the mean LRC for each. We then compute the \textit{changing ratio} of mean LRC values between adjacent segments to gauge irregularity. As depicted in \cref{fig:design:LRC_showcase1}, this variability is especially evident in non-regular runners and during walking. This indicates that a constant LRC ratio cannot be assumed, thus requiring our system to adapt to changing LRC ratios within a single estimation. 
\end{itemize}

We elaborate on our pipeline in the following sections.

\subsubsection{Stride Frequency Estimation.} 
We first conduct footstep detection, which is done using the same approach as heartbeat detection, as discussed in \cref{sec:hrv}, while we filter with a 50Hz low pass filter~\cite{butkow2023heart}. 
By counting the number of detected footsteps, the stride frequency can be estimated.

\subsubsection{Breathing Extraction.} The breathing-related features are then extracted from the in-ear audio.

\textbf{Pre-processing.} 
Human breathing sounds typically fall within the range of 300Hz to 1800Hz \cite{oliveira2014respiratory}. Therefore, a BPF with cutoff frequencies from 300Hz to 1800Hz is used on the input audio during light-intensity rhythmic footstep activities, such as walking. For high-intensity rhythmic footstep activities, where step sounds severely overwhelm the breathing sounds in this frequency range, such as during running, we use a BPF with cutoffs of 2000Hz to 9000Hz to capture harmonics of breathing sounds in these higher frequencies.

\textbf{Breathing template generation.} We generate a breathing template to identify the probability of each frame containing breathing within the estimation window. To generate this template, we collected in-ear audio from a single user while sitting stationary and breathing loudly in a quiet environment. 
We conduct \textit{Pre-processing} and performed \textit{FFT feature generation} on it to generate the breathing template.

\textbf{FFT feature generation:} We divide the audio window into 40ms frames with a 20ms overlap and calculate the periodogram of each frame. Thereafter, we subdivide the breathing frequency range into 15 bins and sum the signal power in each bin from the periodogram. We therefore generate a feature vector with 15 features for each frame, one corresponding to each frequency bin. The breathing template is finally calculated by averaging the feature vectors of all frames. The equally divided frequency scale has better performance for breathing feature extraction due to the low variation of breathing sounds in spectral energy (\cref{s:evaluation}).

\begin{figure}[t]
    \centering
    \includegraphics[width=1\linewidth]{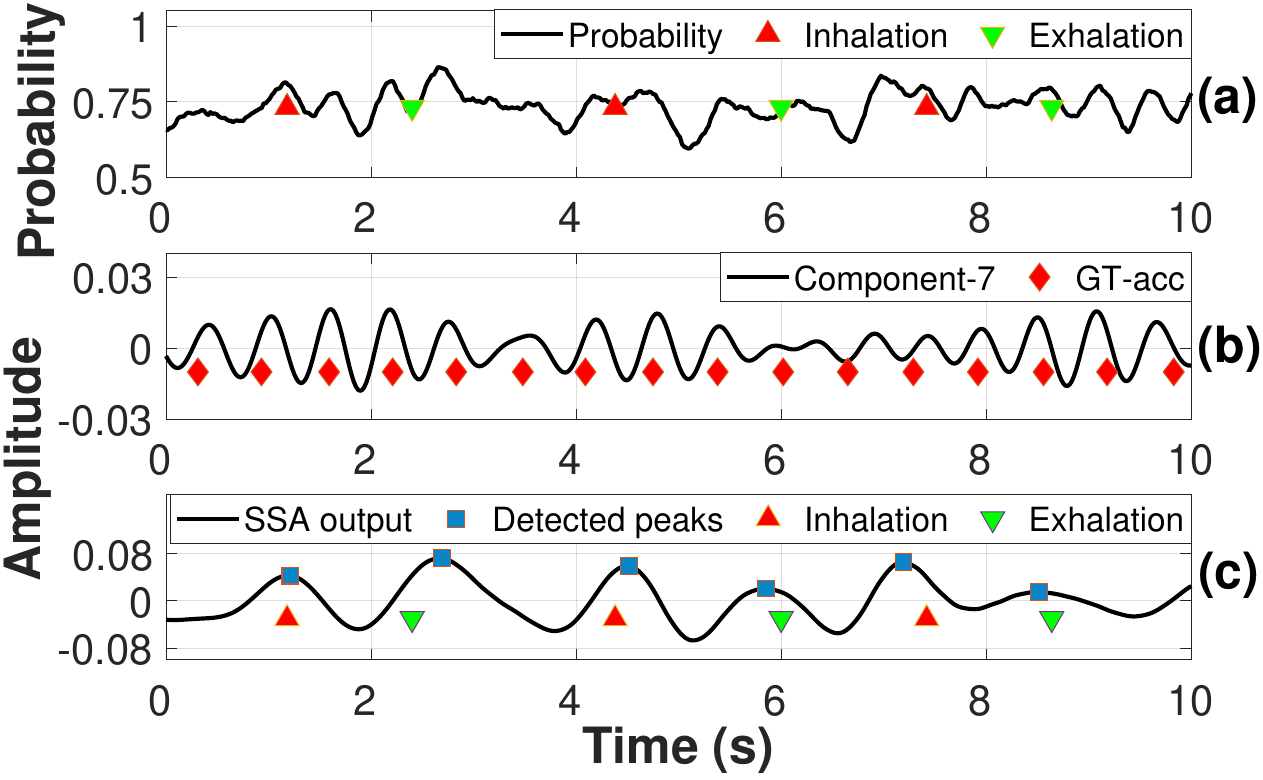} 
    \caption{(a) Probability curve. (b) One component related to step. (c) Extracted breathing pattern with peak detection.}
    \label{fig:design:likelihood_curve}
\end{figure}

\textbf{Probability curve generation.} For each estimation window, we perform the \textit{FFT feature generation} for all frames within it. For each feature vector (\ie corresponding to each frame), we calculate its similarity ($S$) with the breathing template using the cosine similarity~\cite{RunBuddySmartphoneSystemForRunningRhythmMonitoring}. Then, the probability of this frame containing breathing ($P(f)$) is computed by:

\begin{equation}
    P(f)= 
    \begin{cases}
      \frac{S-T}{1-T} & \text{if } S>T\\
      0 & \text{if }S \leq T\\
    \end{cases}    
\end{equation}

where $T$ is a predefined threshold. 
The probabilities from all frames within the estimation window generate a breathing probability curve as shown in \cref{fig:design:likelihood_curve}(a).

\textbf{Probability Curve Decomposition.} Due to the low SNR from strong interference from footstep sounds and light breathing sounds, the breathing pattern is overwhelmed by patterns of interference (\cref{fig:design:likelihood_curve}(a)). To remove the interference patterns, we decompose the probability curve into its constituent components using the SSA algorithm~\cite{elsner1996singular}. SSA is able to effectively separate the underlying components of the curve, allowing for the isolation of periodicity occurring at various time scales, in order of significance, even within highly noisy time series data. 
\cref{fig:design:likelihood_curve}(b) shows component 7 of the decomposed probability curve which clearly corresponds to the steps taken while the user is walking (\ie GT accelerometer data in \cref{fig:design:likelihood_curve}(b) - ``GT-acc''). Once the probability curve is decomposed into periodic components, we select the respiration-related components for RR estimation, and exclude those related to interference.

\subsubsection{Breathing-Related Components Aggregation.} \systemname leverages a loose constraint which can adapt to changes in the LRC ratio to exclude respiration-unrelated components. Specifically, for each decomposed component of the probability curve, we count the number of peaks using peak detection. If the number of peaks falls outside the range of the minimum possible breathing rate ($RR_{min}$) to the maximum possible breathing rate ($RR_{max}$), this component will be regarded as respiration-unrelated and removed. 
$RR_{min}$ can be computed as (similar to $RR_{max}$ with $LRC_{min}$): 
\begin{eqnarray}
 	RR_{min} = ({SF_{est}*({N}/{fs})}) / {LRC_{max}},
\end{eqnarray}
where $SF_{est}$, $fs$ and $N$ are the estimated step frequency, sampling rate, and number of samples in the estimation window, respectively.
$LRC_{max}$ and $LRC_{min}$ are the largest and smallest values of the LRC ratios in humans.

We use the LRC range of 1.9 to 4.9 for light-intensity rhythmic footstep activities~\cite{o2012locomotor}, and 1.8 to 5.6 for high-intensity rhythmic footstep activities~\cite{daley2013impact}. These ranges cover common LRC ratios in humans under each set of scenarios~\cite{o2012locomotor,daley2013impact}, and fully cover the LRC ratios appearing in our collected dataset.
After excluding all breathing-unrelated components, we sum the remaining components into the extracted breathing pattern. Peak detection is then applied to this signal to estimate the final RR (\cref{fig:design:likelihood_curve}(c)).

\subsection{Pipeline Selector}
\label{s:design:svm}

The pipeline selector determines which processing pipeline should be selected for RR estimation, \ie RSA or LRC-based, based on the presence of either clear heartbeat sounds or footsteps in the input signal. If the LRC-based pipeline is selected, we further differentiate this into low-intensity and high-intensity rhythmic footstep activities, so that the correct algorithmic parameters can be applied to the pipeline.

We train our pipeline detector using support vector machines (SVM). 
The in-ear audio is split into 5s segments and Mel-frequency cepstral coefficients (MFCCs) are extracted from each segment and used as the input features to the SVM.
We use a two-stage classifier whereby first we classify a segment as sedentary (\ie strong presence of heart sounds) or active (\ie strong presence of rhythmic footstep sounds). If active, we further classify it into low-intensity and high-intensity rhythmic footstep activities.
Consequently, there are 12 detection results from one model during each 60s estimation window, and we determine the scenario of the whole window through majority voting. 
Specifically, we empirically determine that only consistent results obtained for more than 75\% of segments leads to reliable pipeline selection. Voting aims to handle transition windows between two states that could result in inappropriate pipeline selection. If there is no convergence, the window will be discarded.

%% file: 04-prototyping_data.tex
\section{Implementation and Evaluation}
\label{s:evaluation}

\begin{figure}[t]
    \centering
    \includegraphics[width=1\linewidth]{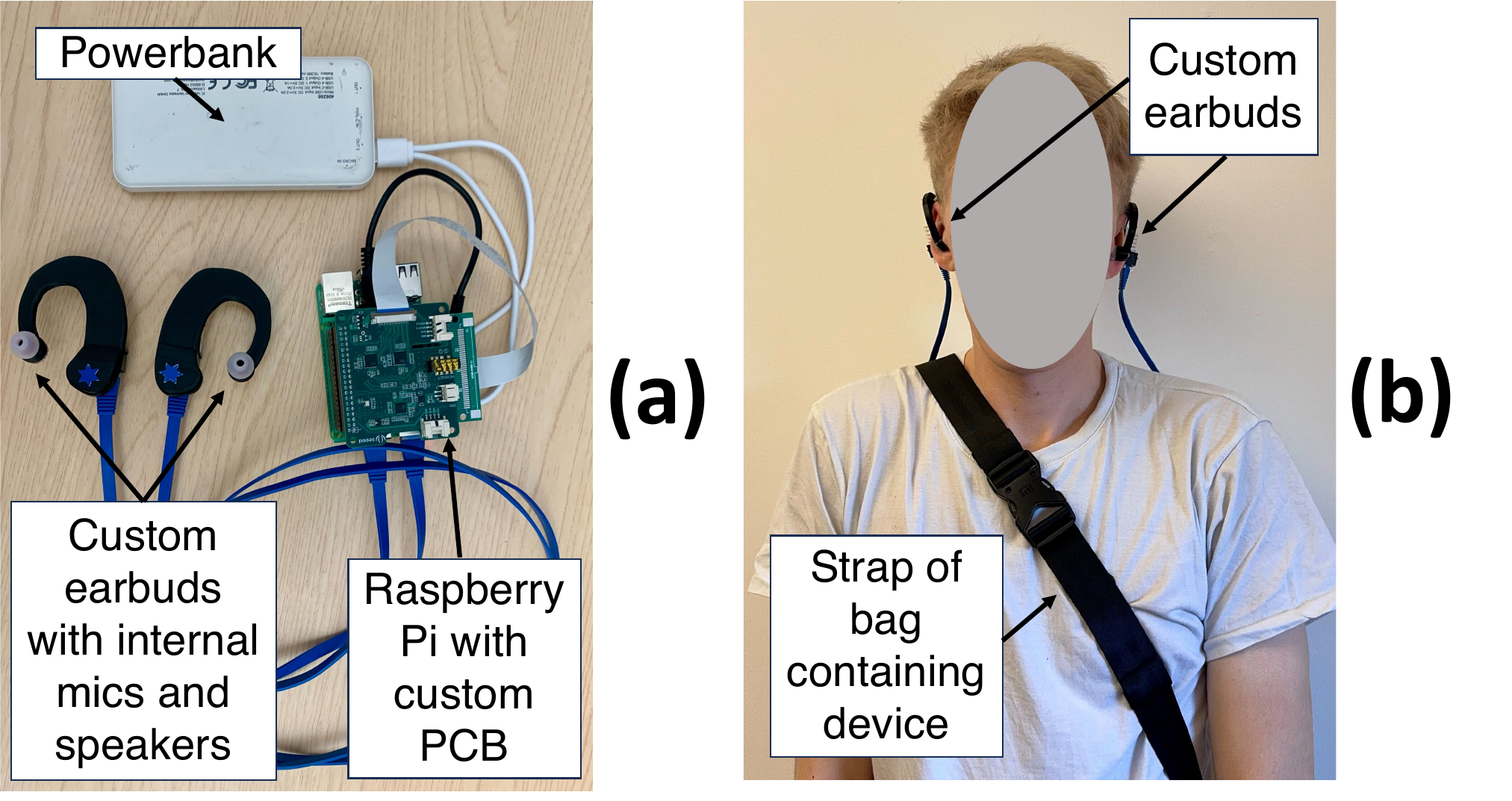}  
    \caption{(a) Custom hardware prototype and (b) one participant wearing the device.}
    \label{fig:prototype}
\end{figure}

\subsection{Implementation}
\subsubsection{Prototyping}
\label{sec:prototype}
Although in-ear microphones are becoming ubiquitous in ANC earbuds~\cite{AirPodsPro2nd, MainFeaturesGalaxy}, no commercial earbud grants access to the raw data. As such, we developed our own custom earbud prototype to collect data. 
We 3D printed an earbud in an ear-hook shape to mount necessary electronic components and allow for a secure attachment. Inside the eartip, we embedded a Knowles SPU1410LR5H-QB microphone~\cite{SPU1410LR5HQB} selected for its flat frequency response between 10~Hz and 10~kHz, thus allowing for low frequency bone conducted sounds to be captured. The microphone was secured inside the earbud facing towards the ear canal. On the left earbud, we embedded a speaker behind the microphone to enable audio playback. 
To capture signals from the sensors, we designed a PCB, which interfaces with an audio codec~\cite{ReSpeaker6MicCircular}  onto a Raspberry Pi 4B. The PCB contains a MCP6004 non-inverting operational amplifier with gain controlled using potentiometers for additional flexibility. The microphone signals are thus amplified before being sampled by the audio codec onto the Raspberry Pi. 
To make the system portable, we placed the Raspberry Pi and PCB into a chest-worn bag and power it with a portable power bank (\cref{fig:prototype}). 

\begin{figure*}
	\begin{minipage}[t]{0.232\linewidth}
            \centering
            \includegraphics[width=1.0\linewidth]{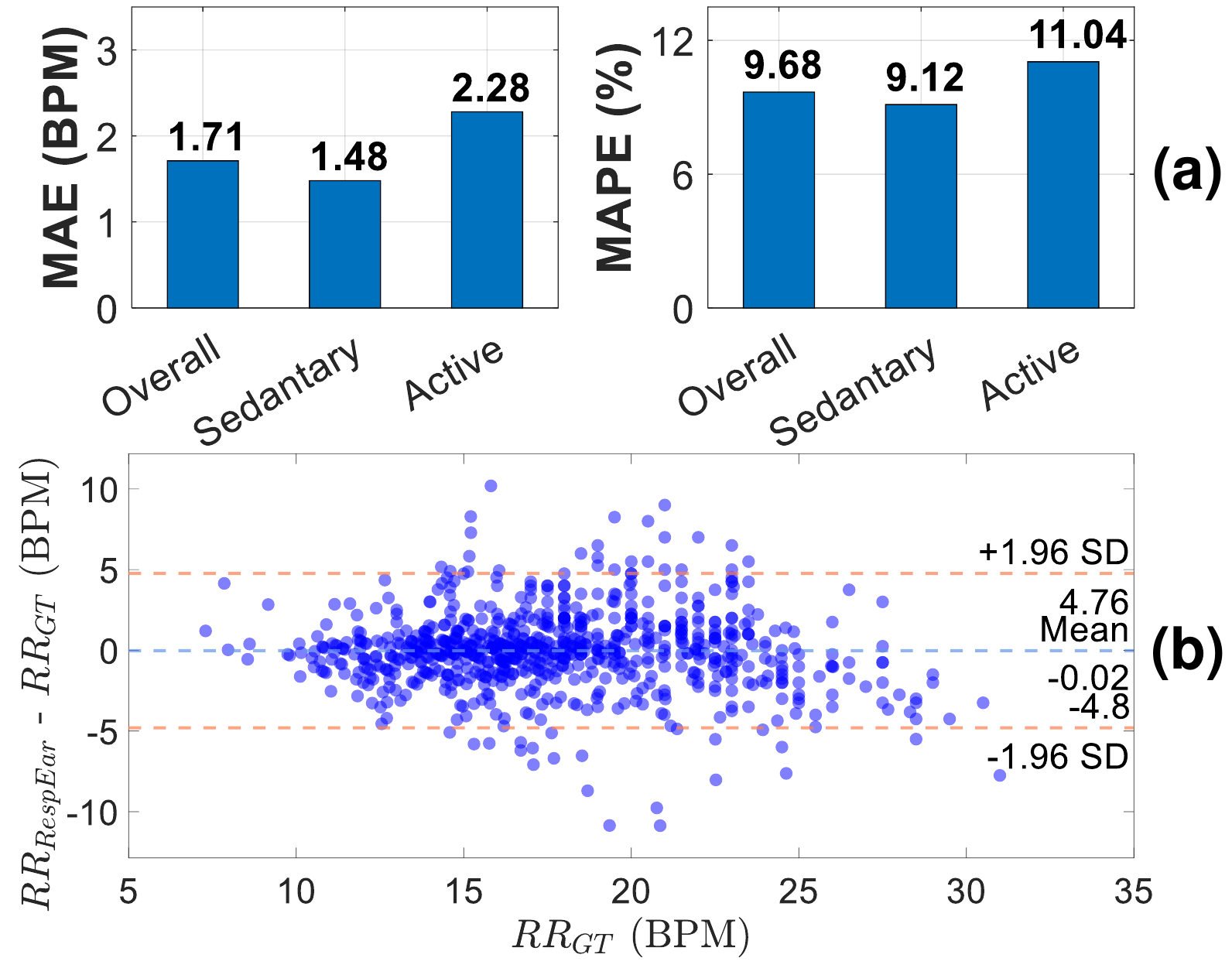}
            \caption{Bar plot of (a) MAE and MAPE. (b) Bland-Altman plot of \systemname.}
            \label{fig:overall_fig1}
	\end{minipage}
	\hspace{0.1cm}
	\begin{minipage}[t]{0.232\linewidth}
    	\centering
    	\includegraphics[width=1.0\linewidth]{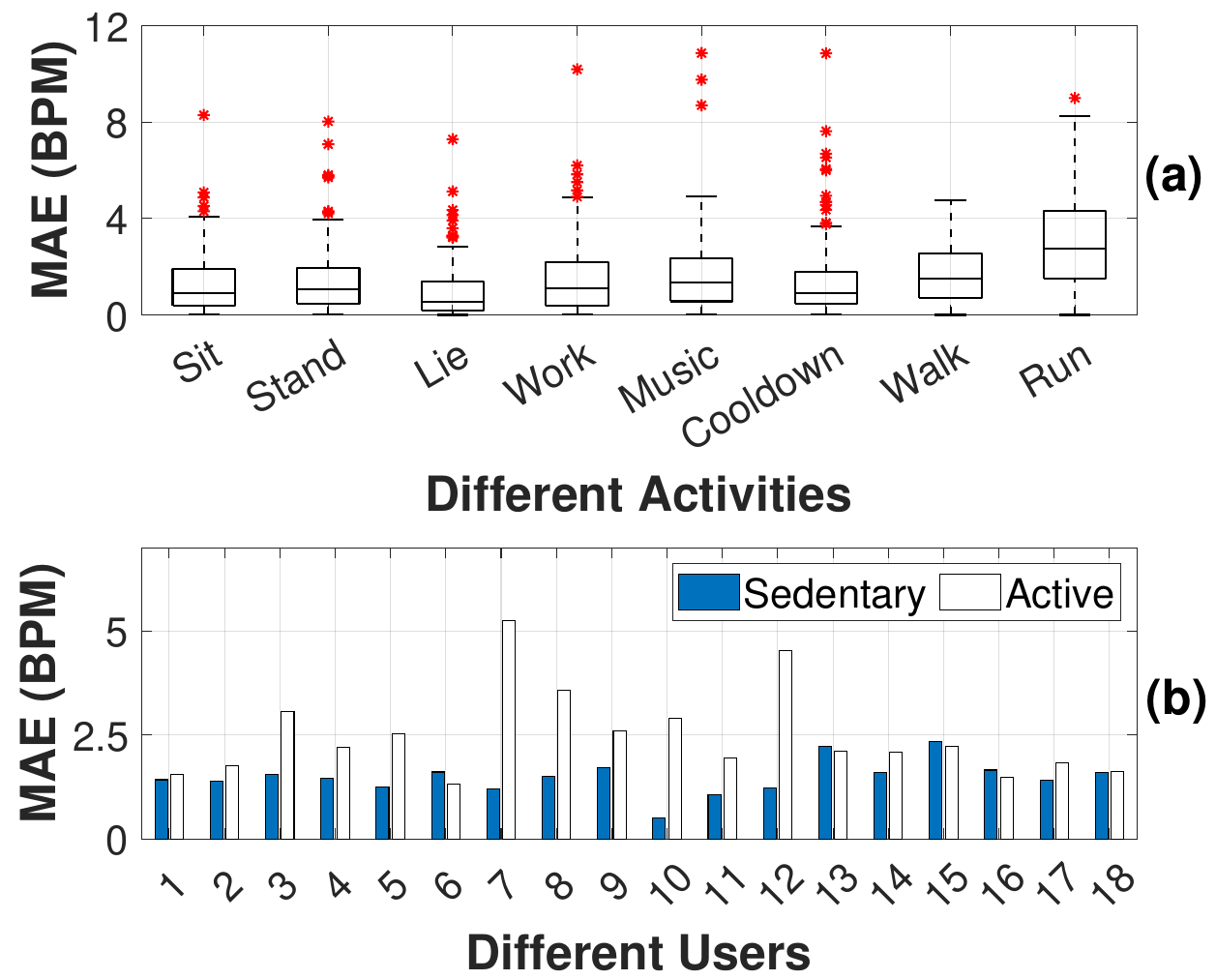}
    	\caption{RR estimation errors for (a) different activities and (b) different users.}
    	\label{fig:acc_multi_atten}
	\end{minipage}
 	\hspace{0.1cm}
 	\begin{minipage}[t]{0.232\linewidth}
            \centering
            \includegraphics[width=1.0\linewidth]{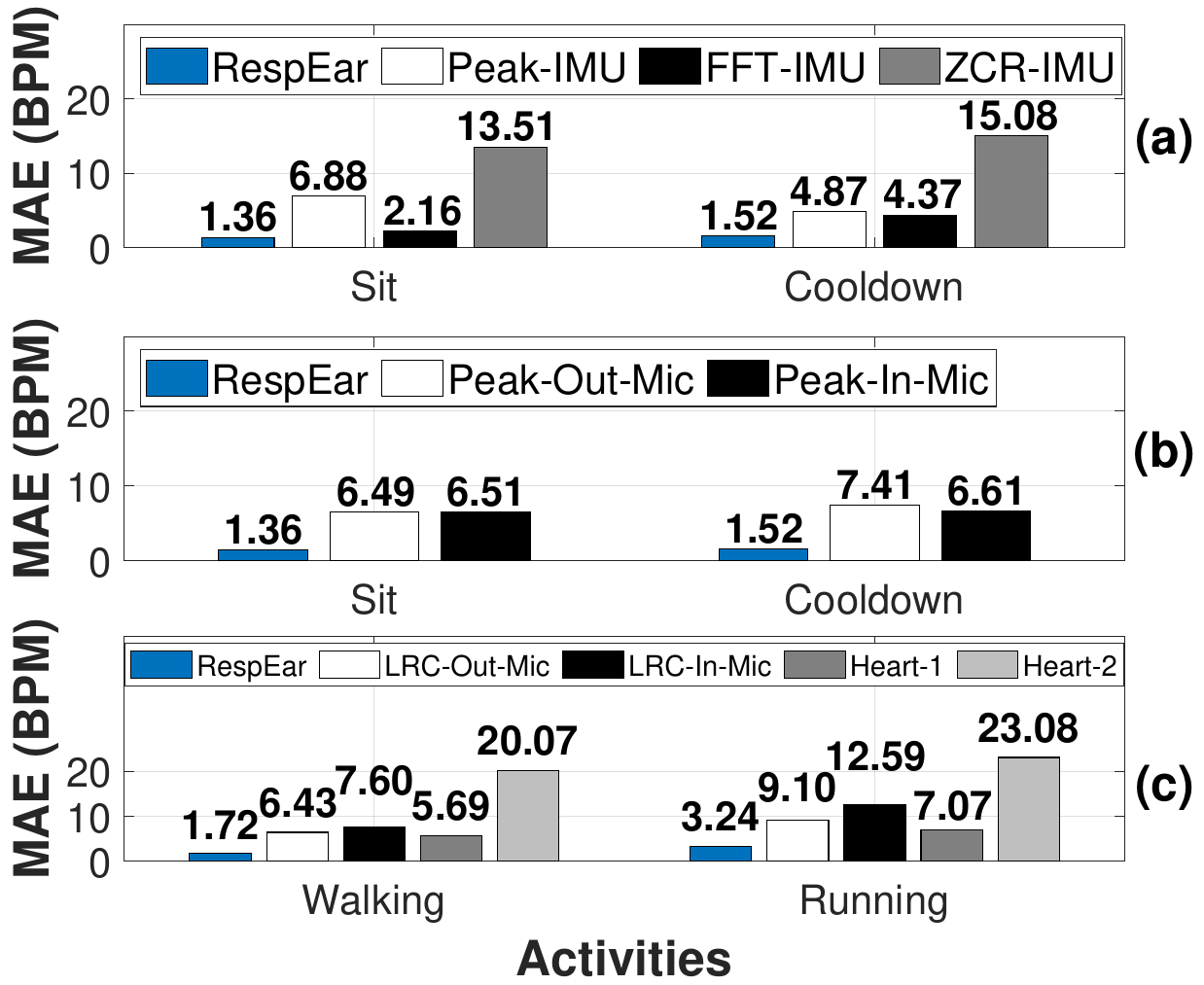}
            \caption{Accuracy comparisons for (a) IMU-based, and (b-c) audio-based approaches.}
            \label{fig:acc_comp}
	\end{minipage}
        \hspace{0.1cm}
 	\begin{minipage}[t]{0.232\linewidth}
            \centering
            \includegraphics[width=1.0\linewidth]{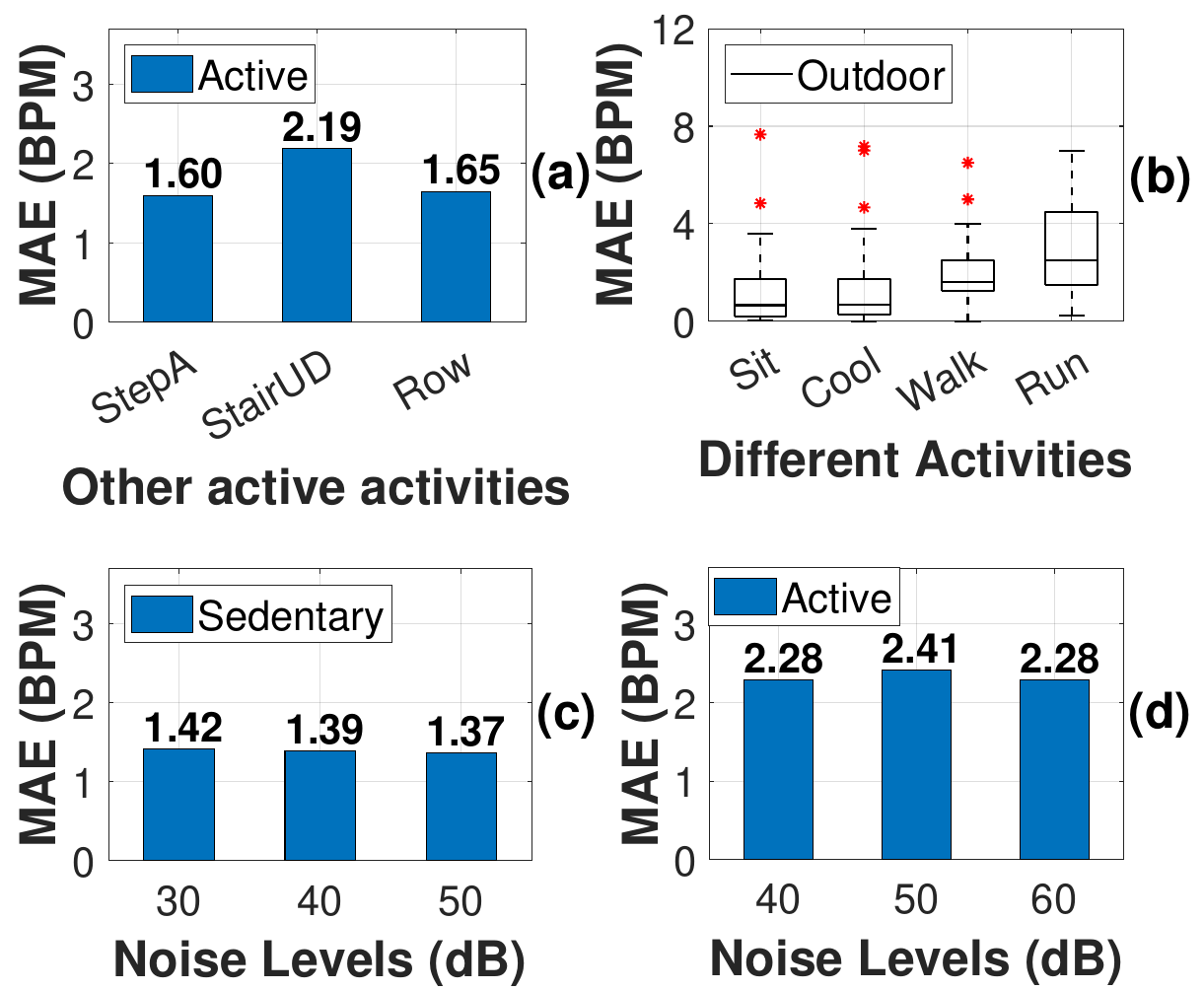}
            \caption{Errors of (a) other rhythmic activities, (b) outdoors, (c-d) noise levels.}
            \label{fig:acc_noise_outdoor}
	\end{minipage}
\end{figure*}

\subsubsection{Data Collection}
\label{section: datacollection}
18 participants (9 male and 9 female) participated in our data collection\footnote{The experiment was approved by the Ethics Committee of the institution.}. The participants' ages ranged between 22 and 51. The participants underwent sedentary and active activities, with each activity was performed for 5 minutes. The sedentary activities are: (1) sitting; (2) standing; (3) lying down; (4) listening to music (performed for the duration of one song); (5) working in the wild; (6) uncontrolled cooldown after exercise. The active activities are (7) walking; and (8) running. 
The activities encompass typical scenarios when a person uses earables and requires RR monitoring. Our data collection involved controlled activities and also in-the-wild scenarios to ensure the applicability of our methods to real-world use. 
No breathing rates were imposed, and participants were free to breathe as they wished.
Breathing after exercise was performed immediately after the user had completed their running to capture their natural cool-down breathing. While sitting and standing, users were asked to move their heads three times to capture head motions to assess the impact of our interference artefact filtering algorithm. 
Active activities were performed on a treadmill and participants chose their comfortable paces to walk/run. We use the Zephyr BioHarness 3.0 chest strap~\cite{Zephyr} to collect GT RR with 25~Hz sampling rate and the audio data from in-ear microphone is collected at 22050~Hz.

%% file: 05-evaluation.tex
\subsection{Evaluation}
\subsubsection{Metrics}

We evaluate system performance using the Mean Absolute Error (MAE)~\cite{InEarAudioWearableMeasurementOfHeartAnda,RRMonitorResourceAwareEndtoEndSystemForContinuousMonitoring} which is the average absolute error between the GT RR and the calculated RR for each window. We also use the Mean Average Percentage Error (MAPE), the average percentage absolute error. 
\subsubsection{\systemname Overall Performance}

\textbf{Overall performance.} We present the overall performance of \systemname in \cref{fig:overall_fig1}(a). \systemname achieves an overall MAE of 1.71BPM (MAPE of 9.68\%), with a MAE of 1.48BPM (9.12\%) and 2.28BPM (11.04\%) for sedentary and active respectively. The Bland-Altman plot for \systemname is provided in \cref{fig:overall_fig1}(b). We achieve a very low mean error of -0.02BPM with narrow limits of agreement of -4.8 to 4.76. This indicates very good agreement between \systemname and ground truth breathing rate measurements, highlighting the strength of our system. 

\textbf{Performance Per Activity. }
\cref{fig:acc_multi_atten}(a) provides a boxplot of the overall performance of \systemname for each activity. The performance of each sedentary activities are comparable. Slightly higher errors exist while listening to music (MAE=1.98BPM) (detailed in \cref{sec:music}), and working (1.56BPM). This is because working is an uncontrolled activity and thus participants were more active during this task, leading to more interference artefact. The estimation errors while walking and running are satisfactory, \ie walking (MAE = 1.75BPM; MAPE = 9.17\%), and running (MAE = 3.12BPM; MAPE = 14.01\%). The slightly higher running errors are due to the higher footstep interference.

\textbf{Individualised Performance. }
\cref{fig:acc_multi_atten}(b) reports the overall performance of \systemname for each participant.  
It is evident that the MAE while sedentary is consistent amongst users, with no user exceeding 2.3BPM error. There is much more variation amongst estimation errors while active: the smallest MAE is 1.26BPM for user 16, with the largest MAE being 5.25BPM for user 7. 
The majority of error comes from 2 users, \ie user 7 and 12. User 7 ran at 5KPH, which has slightly worse performance than faster running speeds (discussed in \cref{sec:speed}). User 12's running generated large noise because the feet kept hitting the side of the treadmill, resulting in high-energy noise across all frequencies in the in-ear audio. However, regardless of this, the system still generalizes well for the majority of users over all activities.

\subsubsection{Baseline Comparison. }\label{sec:baseline}
We compare the performance of our system to those of existing works for each of the three sensors mentioned in \cref{sec:initial_exploration}. To perform this study, we collected data from 11 users in a combination of indoor and outdoor settings, while the subjects were sitting still, cooling down after running (sedentary), walking and running (active). 
We implemented the three IMU based algorithms for RR estimation under sedentary conditions in~\cite{RRMonitorResourceAwareEndtoEndSystemForContinuousMonitoring}: an FFT approach (FFT), a peak detection approach (Peak) and a zero-crossing rate (ZCR) approach (\cref{fig:acc_comp}(a)). For the in-ear and out-ear microphones, we implemented the algorithm for sedentary estimation in~\cite{InEarAudioWearableMeasurementOfHeartAnda} which uses a peak detection approach where peaks are detected from the envelope of the microphone signal (\cref{fig:acc_comp}(b)). 
For active, we implemented the algorithm employed by \cite{RunBuddySmartphoneSystemForRunningRhythmMonitoring,gu2017detecting} (LRC), and expanded upon it to calculated RR (\cref{fig:acc_comp}(c)). 
From \cref{fig:acc_comp} , it is evident that under both sedentary and active scenarios, our system significantly outperforms the methods in the literature, thus highlighting both the strength of the in-ear microphone and our processing pipeline. 

We reproduced two motion-resilient HR estimation methods \cite{butkow2023heart,EarmonitorInearMotionresilientAcousticSensingUsingCommodity} to assess their applicability in \systemname while active. Specifically, we derived HRV from the generated ``ECG'' signal by \cite{butkow2023heart} during walking and running, integrating it into our RSA-based RR monitoring pipeline (Heart-1). Following \cite{EarmonitorInearMotionresilientAcousticSensingUsingCommodity}, we eliminated walking/running frequency components from in-ear audio and then applied a [0.1, 0.8]Hz band-pass filter for RR estimation (Heart-2). \cref{fig:acc_comp}(c) shows that these methods cannot be applied for RR monitoring while active, necessitating our LRC-based pipeline.

\subsubsection{Benchmark Evaluations}
\begin{figure*}
  	\begin{minipage}[t]{0.232\linewidth}
		\centering
		\includegraphics[width=1.0\linewidth]{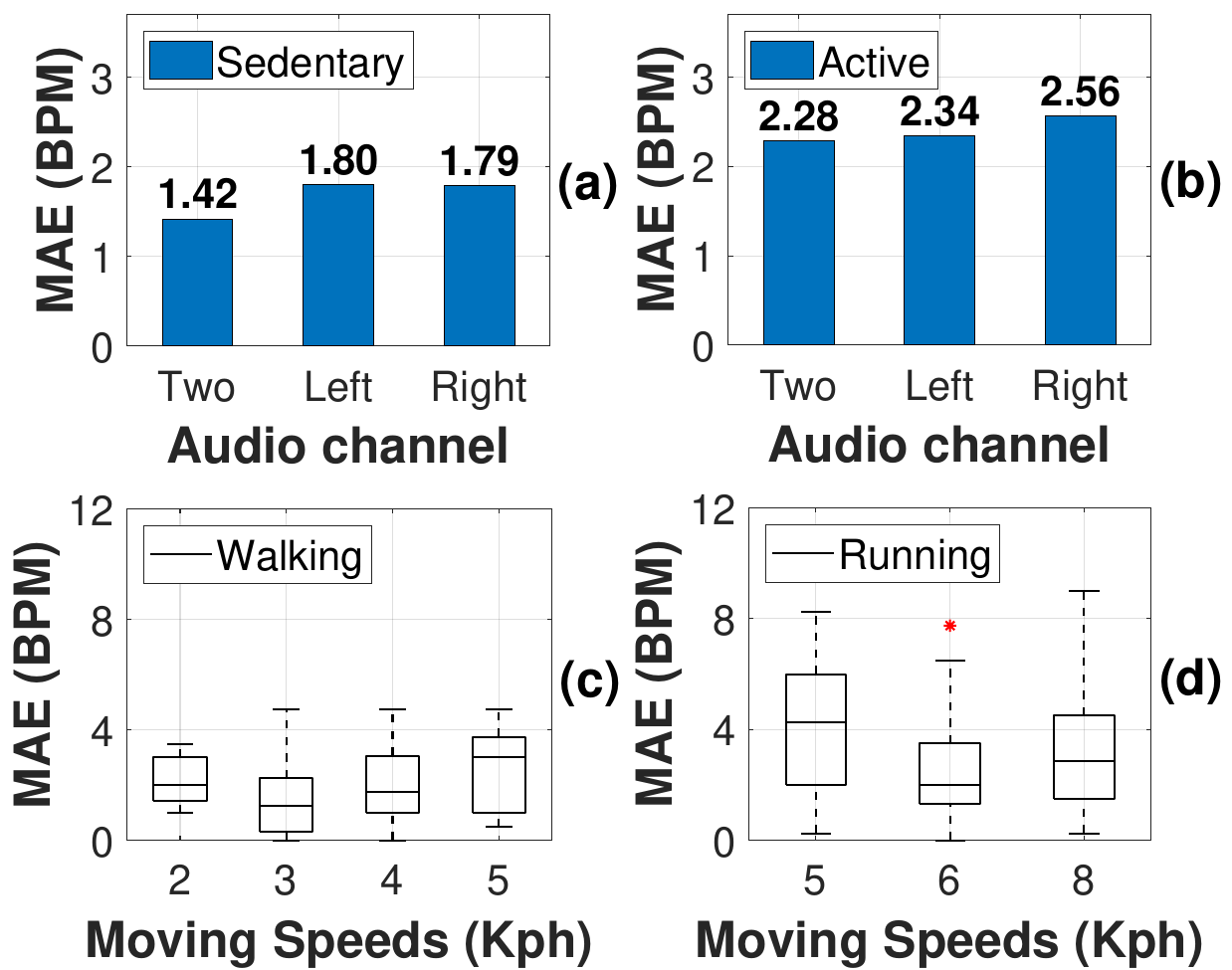}
		\caption{Estimation errors for different (a-b)  audio channels and (c-d) moving speeds.}
		\label{fig:diff_audio_diff_speeds}
	\end{minipage}
	\hspace{0.1cm}
        \begin{minipage}[t]{0.232\linewidth}
		\centering
		\includegraphics[width=1.0\linewidth]{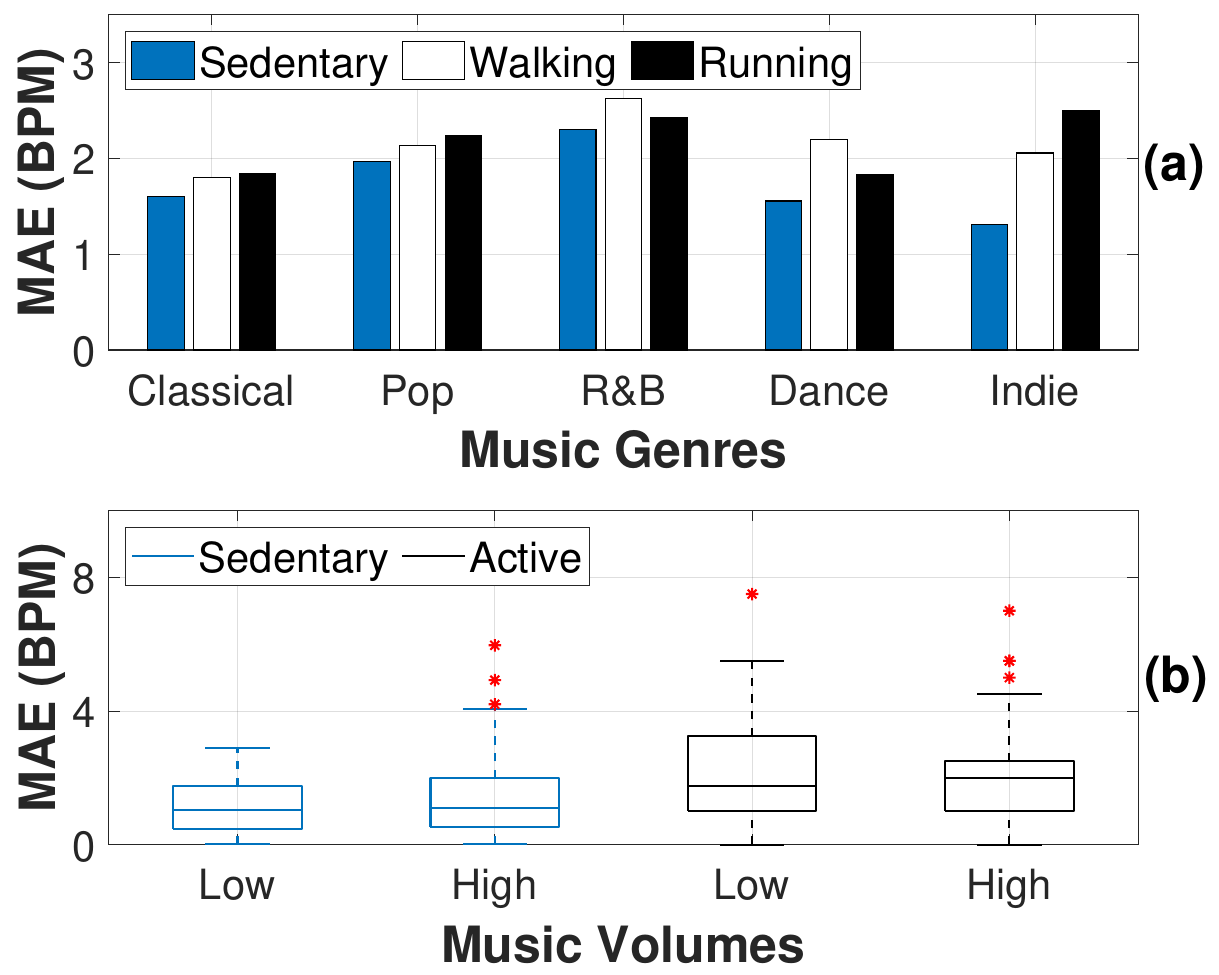}
		\caption{Errors with music: (a) across  genres and (b) at different volume levels.}
		\label{fig:music_comparison}
	\end{minipage}
        \hspace{0.1cm}
 	\begin{minipage}[t]{0.232\linewidth}
    	\centering
    	\includegraphics[width=1.0\linewidth]{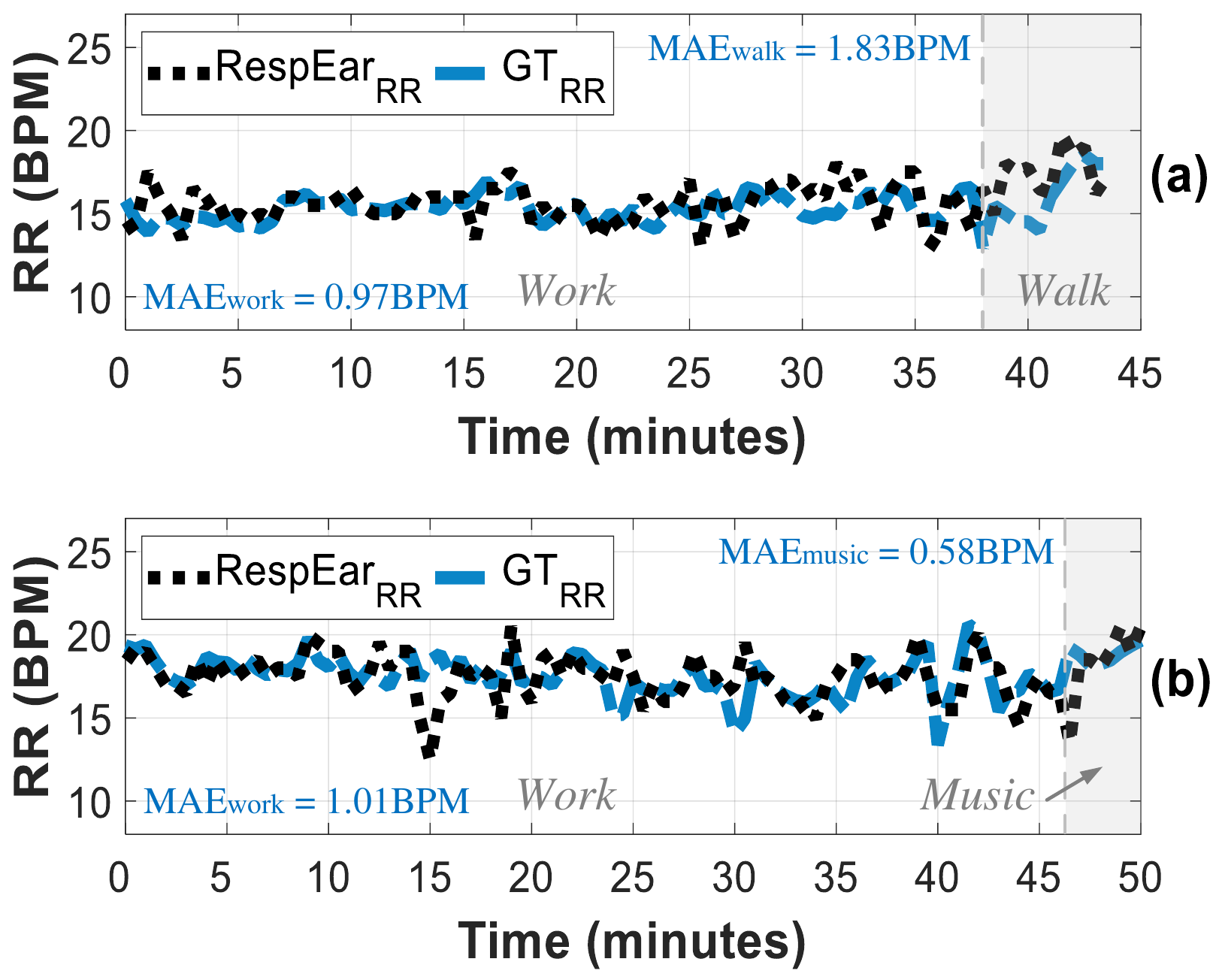}
    	\caption{Errors of longitudinal in-the-wild tracking for (a) user A and (b) user B.}
    	\label{fig:acc_in_the_wild}
	\end{minipage}
        \hspace{0.1cm}
	\begin{minipage}[t]{0.232\linewidth}
    	\centering
      	\includegraphics[width=1\linewidth]{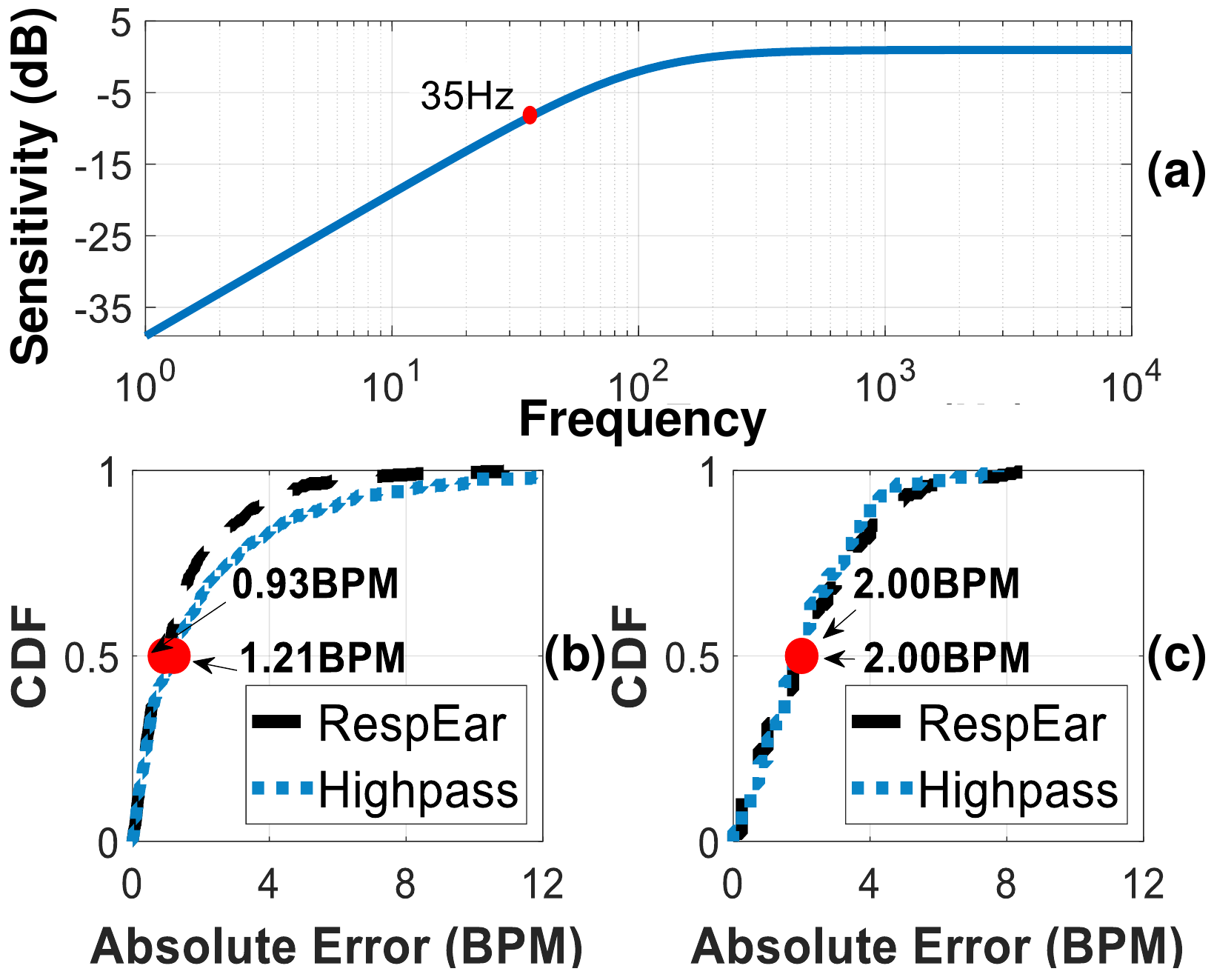}  
            \caption{Errors with (a) high-pass filter on (b) sedentary and (c) active scenarios.}
            \label{fig:apg_justification}
	\end{minipage}
\end{figure*}

\textbf{Other Active Activities.} We assessed \systemname's performance when users engaged in other activities of varying intensities involving rhythmic footsteps, including step aerobics (StepA), climbing up and down stairs (StairUD), and rowing on an indoor rower (Row). We recorded 5 minutes of data per activity per participant, with activities performed in an uncontrolled manner. \cref{fig:acc_noise_outdoor}(a) indicates \systemname works properly during different rhythmic activities using the LRC-based pipeline, demonstrating \systemname's effectiveness for activities with rhythmic footsteps.

\textbf{Outdoor Performance.} 
We also assessed performance outdoors in an uncontrolled environment. The tests were performed on a concrete pavement outside an academic building next to a building site with active construction happening. We assessed the performance under different activities, including sitting, walking, running and cooling down. We recorded 5 minutes of data per activity for each participant. 
When walking and running, participants were free to select their preferred pace and move around the area. There were thus natural changes in pace throughout the experiment to test whether our system functions under both controlled and uncontrolled speeds. 
\cref{fig:acc_noise_outdoor}(b) shows the results, indicating \systemname achieves robust performance outdoors.

\textbf{Impact of Ambient Noise Level. }
Since \systemname is audio-based, it is essential to ensure that it functions as expected in the presence of ambient noise at different levels. We assess this performance in \cref{fig:acc_noise_outdoor}(c) and (d). We see that over three ambient noise levels for sedentary (\cref{fig:acc_noise_outdoor}(c)) and active (\cref{fig:acc_noise_outdoor}(d)), \systemname achieves consistent results. This is due to occluding the ear canal using earbuds which attenuates ambient noise and the occlusion effect which attenuates high frequency external noise and amplifies low frequency heartbeat and footstep sounds~\cite{OESenseEmployingOcclusionEffectForInearHuman, butkow2023heart}. 

\textbf{Impact of Audio Channel.}
\cref{fig:diff_audio_diff_speeds}(a) provides the overall MAE in the left and right channels and the channel chosen by \textit{automatic channel selection} (denoted by \textit{two}) while sedentary. The individual performance of the two channels is similar with a MAE of 1.8BPM and 1.79BPM for left and right respectively.
After the \textit{auto channel selection}, the MAE is significantly reduced (1.42BPM), indicating the efficacy of our design. 
Moreover, even from one channel, the performance is still good, proving that \systemname can be used even when the user is wearing a single earbud. \cref{fig:diff_audio_diff_speeds}(b) provides the MAE for the left, right and fusion of two channels (\ie we use the mean value of the estimated RR from two channels) while active.  Again, we see that the fusion channel has a lower error than that of the two channels, with a MAE of 2.28BPM for the fusion compared to 2.34BPM and 2.56BPM for left and right respectively.

\textbf{Impact of Moving Speed. }\label{sec:speed}
In \cref{fig:diff_audio_diff_speeds}(c) and (d), we assess the impact of different speeds while active. 
For both walking and running, the median error remains similar regardless of speeds. We see the slightly higher error under running at 5km/h. 
This may be attributed to the lower SNR of low-speed running, which causes similar interference from running footsteps but induces weaker breathing sounds compared to higher speeds. 
However, even with this slight trend, \systemname still achieve good results across different speeds.

\textbf{Impact of Music Listening.}
\label{sec:music}
We assessed the impact of listening to music on earables while users were sedentary, walking and running. For each condition, we played music on the left channel covering a range of common genres with different volumes.
\cref{fig:music_comparison}(a) and \cref{fig:music_comparison}(b) provides a comparison of performance per genre per activity and performance with different music volumes respectively.
We see the performance across all genres and with different playing volumes while sedentary is comparable with the sedentary performance without music playing (the error of left channel as shown in \cref{fig:diff_audio_diff_speeds}(a), \ie 1.8BPM). This is due to the minimal overlap of heart sound frequencies with the music frequency, with only 0.5\% of music energy lying in the 0-30Hz range. While active, the performance across all genres is slightly higher than the one without music. Similarly, errors are 2.50BPM and 2.92BPM for soft and loud music respectively compared to 2.34BPM when active without music (the error of the left channel in \cref{fig:diff_audio_diff_speeds}(a)). We see that again, music does not significantly degrade RR estimation performance while active. There is again a non overlap between music and stride frequency, meaning that music does not impact stride detection accuracy. Although music and breathing sounds have overlapping frequencies, we do not use breathing sounds directly, but rather generate a breathing template to detect the probability of breathing sound. This means that breathing and music can easily be differentiated, mitigating the impact of music of accuracy.

\textbf{In-the-Wild Performance.}
We asked users to wear the device for an hour in a busy office while undergoing standard daily activities. 
The users worked at their desks, listened to music, walked around the office, and performed other activities as they wished, such as sipping coffee or using their cell phones. \cref{fig:acc_in_the_wild} shows the tracking performance for two users, \ie continuously estimated RR from \systemname (one RR estimation per 30s) compared with $GT_{RR}$. User A (shown in \cref{fig:acc_in_the_wild}(a)) worked and walked around the office. User B (\cref{fig:acc_in_the_wild}(b)) worked at their desk and  listened to music while working. User A achieved a MAE of 0.97BPM and 1.83BPM while working and walking respectively, and User B achieved errors of 1.01BPM and 0.58BPM while working and listening to music respectively. 
These errors are consistent with the results obtained in the laboratory study for these two users, 
proving that \systemname has excellent performance both in controlled laboratory settings and uncontrolled, real-world settings. It is also clear that \systemname can accurately track RR longitudinally, even in an uncontrolled setting.

\textbf{Impact of ANC.}
Recently, there has been an increase in literature on using active sensing for physiological sensing~\cite{APGAudioplethysmographyForCardiacMonitoringInHearables} due to concerns that passive sensing may not always function under the influence of ANC. This is on account of high pass filters in ANC earables that attenuate low frequency sounds (less than 50Hz~\cite{APGAudioplethysmographyForCardiacMonitoringInHearables}). To assess this, we apply a high-pass filter to our data (with the low cutoff mimicking the hardware filters on a commercial ANC system~\cite{fan2023design}) and run our processing pipelines. The results of this study are provided in \cref{fig:apg_justification}. Under both conditions, the median error is mostly unaffected by the filter, with the slightly higher error in sedentary due to the very low frequency of heartbeat sounds. However, overall, this study proves that \systemname will function accurately on ANC earables. Concretely, although we are filtering out the signal energy of heartbeat sounds, due to the high quality occlusion, we also get signal components after attenuation on low-frequency sounds which can be reliably used for estimation. Further, if ANC filters are implemented using software~\cite{APGAudioplethysmographyForCardiacMonitoringInHearables}, filtering can be applied selectively based on the application.

\subsubsection{System Components Evaluation}

\begin{figure}[t]
	\begin{minipage}[t]{0.48\linewidth}
		\centering
		\includegraphics[width=1.0\linewidth]{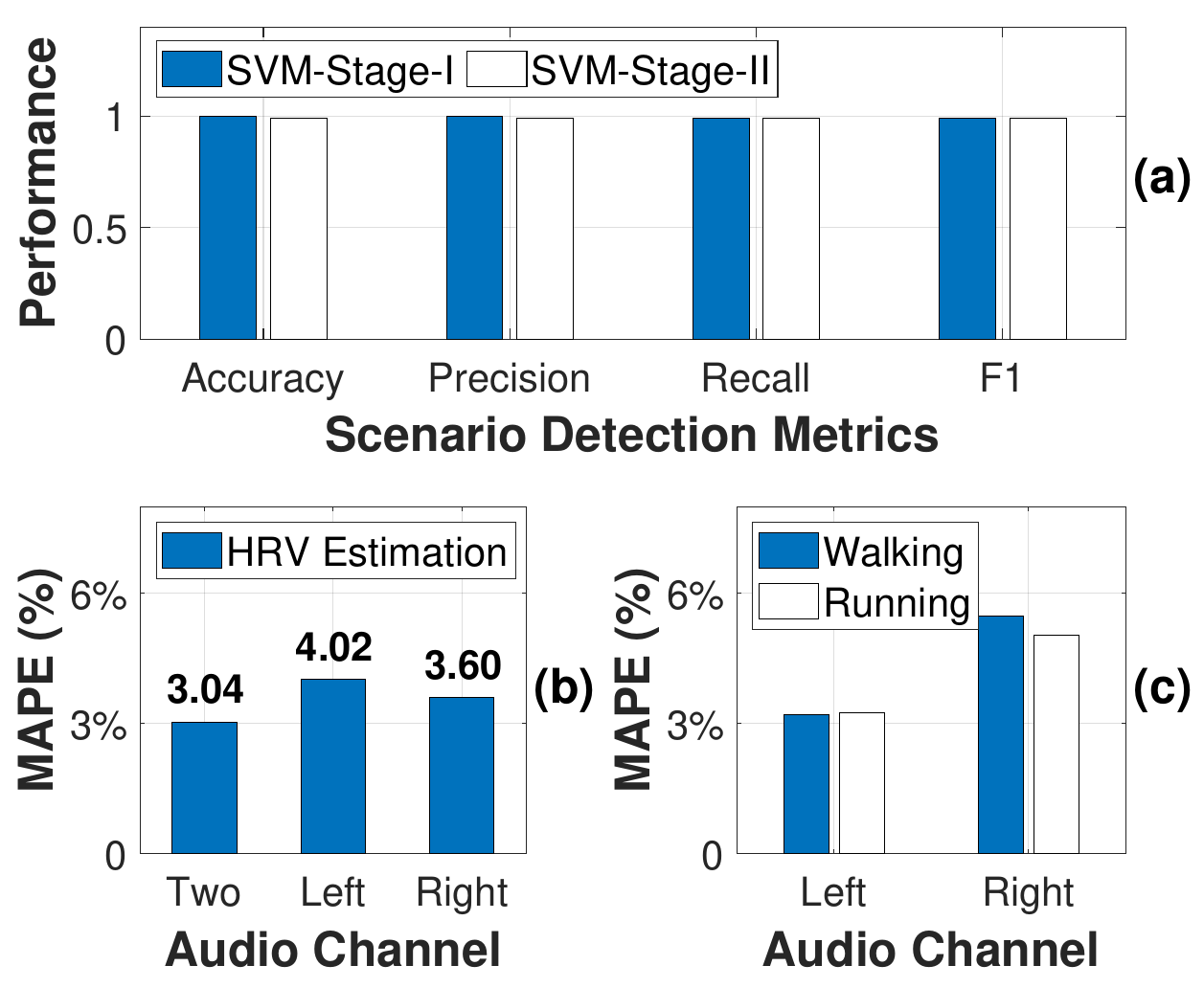}
		\caption{Performance of system components: (a) Pipeline selector, (b) HRV, and (c) Stride frequency.}
		\label{fig:acc_components}
	\end{minipage}
	\hspace{0.1cm}
	\begin{minipage}[t]{0.48\linewidth}
    	\centering
    	\includegraphics[width=1.0\linewidth]{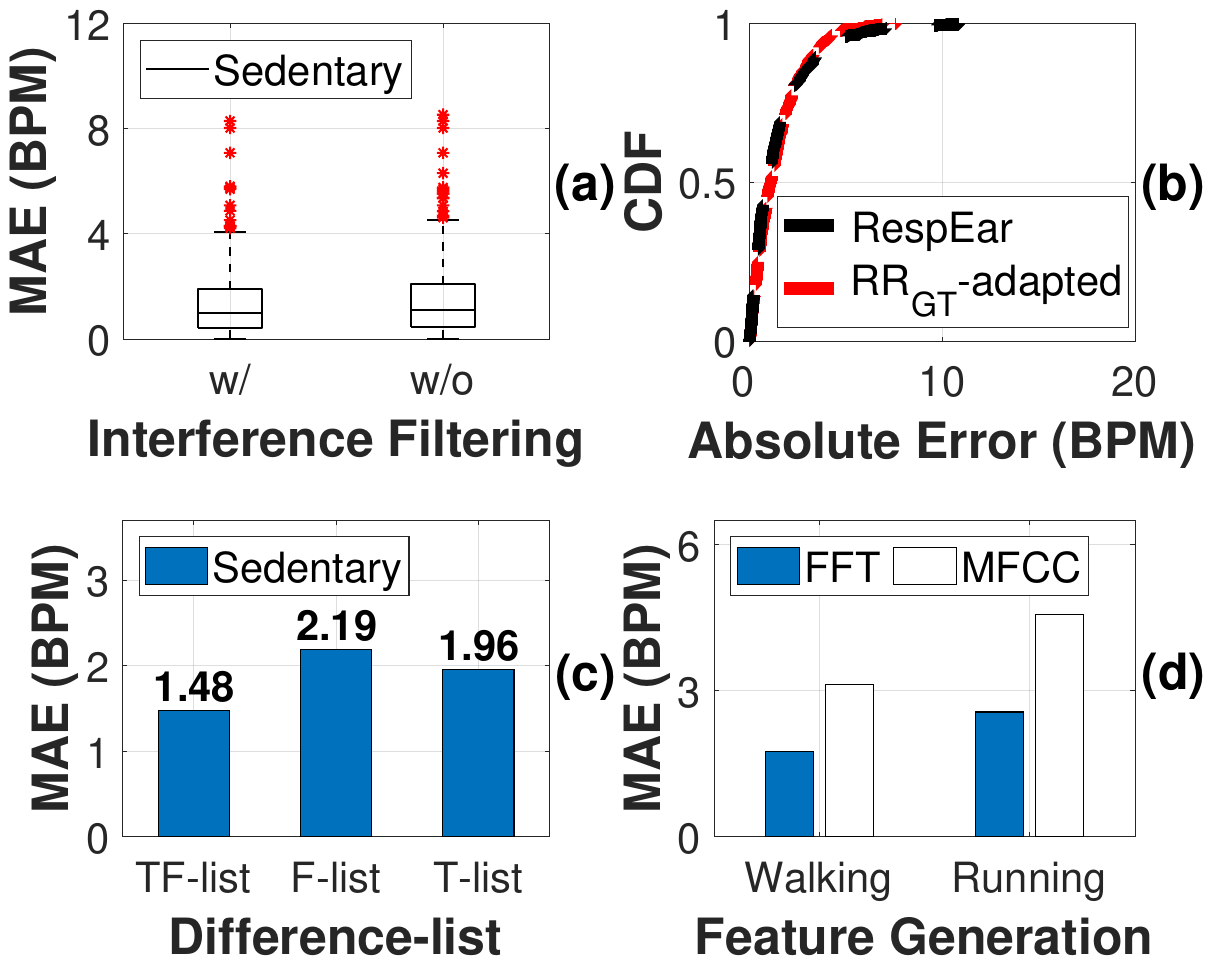}
    	\caption{Impact of (a) interference filtering, (b) adaptive BPF, (c) difference list, (d) FFT features.}
    	\label{fig:ablation}
	\end{minipage}
\end{figure}

\textbf{Pipeline Selector.}
\cref{fig:acc_components}(a) provides the results of our pipeline selector module using SVM on 5s segments. The SVM is trained on 13 users' data (randomly chosen during training) and tested on the remaining 5 users to ensure user independence of the train and test sets. We implemented 5-fold cross validation and report the average results over 5 folds. Our system is able to select pipelines with excellent performance, achieving 100\% accuracy for determining whether a window is active or sedentary (SVM-Stage-I), and 99\% accuracy for determining whether an active window is low-intensity or high-intensity rhythmic footstep activities (SVM-Stage-II). With majority voting of results across an estimation window, the detection accuracy on both tasks is $100\%$.

\textbf{Accuracy of HRV Estimation and Stride Detection.}
To assess the accuracy of our HRV estimation, we compute the MAPE between the GT HRV from ECG and our estimated HRV on beat-to-beat level (\cref{fig:acc_components}(b)), where we compare performance on the left channel, right channel and \textit{automatic channel selection}. Using \textit{automatic channel selection}, we achieve the best performance with a MAPE of 3\%, which is competitive with reported results for ECG on well-known datasets, \eg the MIT-BIH Arrhythmia Database~\cite{arefeen2021interbeat}. Our results for stride frequency estimation are provided in \cref{fig:acc_components}(c). Our system detects strides with a MAPE of less than 3\% for both walking and running using channel fusion, again competitive with literature on in-ear step counting~\cite{OESenseEmployingOcclusionEffectForInearHuman}.

\textbf{Ablation Study.}
\textbf{1) Interference artefact filtering:} First, in \cref{fig:ablation}(a), we examine the impact of our interference artefact filtering. The MAE without interference artefact filtering is 1.63BPM, and that with the filter is 1.45BPM, showing that we get a 12\% decrease in error.
We also see that interference artefact filtering decreases variance and median error in estimation, making estimations more robust. \textbf{2) Adaptive breathing signal extraction:} \cref{fig:ablation}(b) illustrates the performance of \systemname while sedentary compared to the performance obtained when directly using the adaptive filter generated from $GT_{RR}$ (detailed in \cref{s:design:stationary:prin}). \systemname achieve a MAE of 1.48BPM, and with the adaptive filter from the $GT_{RR}$, we obtain an error of 1.44BPM. We thus see that the proposed design can lead to estimations very close to the upper bound of performance using the RSA-based RR estimation principle (noting $GT_{RR}$ is unknown in practical estimation). \textbf{3) Calibration from T-difference-list:} \cref{fig:ablation}(c) shows the importance of the fusion of the T-Difference-List and the F-Difference-List for achieving accurate estimations, with the fusion technique lowering error by 32.43\% over the F-Difference-List. \textbf{4) FFT feature generation:} In \cref{fig:ablation}(d), we compare using our FFT features to MFCC features for creating the breathing features for RR estimation under activity. Although MFCCs are most commonly used for audio feature extraction~\cite{SHARMA2020107020}, we gain a significant performance increase for both walking and running. \textbf{5) Impact of LRC range selection:} \systemname leverages the common LRC ranges for walking and running in humans respectively. To further test its robustness, we test \systemname while active using an increased LRC range for walking and running, \ie $\pm$0.2. The results show that the performance is comparable, \ie 0.2BPM difference for walking and running.

\subsubsection{System Overhead on Smartphone}

For portable solution, we deployed \systemname prototype on an iPhone 12 Pro with 8GB of memory, and a 2477~mAH battery capacity (2815~mAH battery with 88\% battery health) to measure system overhead. 
\systemname's latency is 3.11s per window while sedentary and 12.27s per window while active. Since processing occurs in 60s windows, with a 30s overlap, a new RR estimate can be made every 30s under both conditions, implying that our system can run in real time. We have not considered possible data transfer costs via radio frequency (\eg BLE) which would add additional small delays depending on scenarios. When run continuously for an hour, \systemname consumed 4\% and 14\% battery for the RSA-based and LRC-based pipelines, respectively. The LRC-based pipeline consumes more power due to the longer latency on account of the more complex algorithm. To contextualise this, if playing music for an hour, the same phone battery decreased by 8\%, showing that our application lies within standard levels for battery consumption: since people spends 8 hours on average of their day performing sedentary activities~\cite{SedentaryLifestyleOverviewOfUpdatedEvidenceOf} and lie down all night, the impact on battery life will, on average, be minimal. Further standard processing optimizations could be applied. 
In terms of memory, \systemname consumes a maximum of 49.1MB and 49.3MB per window equating to 0.3\% of the available memory on the device. Overall, we see that \systemname can feasibly be run on a smartphone longitudinally to potentially provide real-time RR monitoring.

%% file: 07-related_work.tex
\section{Related Work}
\label{s:related}

\subsection{Non-earable based RR monitoring}

\textbf{Smartwatch.} \textbf{IMUs:} \cite{BioWatchEstimationOfHeartAndBreathingRates, SleepMonitorMonitoringRespiratoryRateAndBodyPositiona, WearBreathingRealWorldRespiratoryRateMonitoringUsing, hao2017mindfulwatch} use IMUs on smartwatches for RR monitoring. Most~\cite{BioWatchEstimationOfHeartAndBreathingRates,SleepMonitorMonitoringRespiratoryRateAndBodyPositiona,hao2017mindfulwatch} aim to estimate RR under stationary conditions. 
\cite{WearBreathingRealWorldRespiratoryRateMonitoringUsing} demonstrates the possibility of working under both stationary and walking conditions using learning-based techniques. However, it detects and rejects sensor data that are unsuitable for RR extraction. For walking, no more than 20\% of the data windows are accepted. Some commercial smartwatches~\cite{Garmin,AppleWatch} have integrated the RR estimation function, which works effectively only at rest. 
\textbf{PPG:} \cite{RobustRespiratoryRateMonitoringUsingSmartwatchPhotoplethysmography,dai2021respwatch} perform RR estimation on smartwatches using PPG. \cite{RobustRespiratoryRateMonitoringUsingSmartwatchPhotoplethysmography} employs learning-based solutions, working under both sedentary and moving conditions, but it struggles to provide reliable estimations during moving activities (MAE = 3.94 BPM). \cite{dai2021respwatch} utilizes both training-free and learning-based methods, tailored for scenarios where the user is engaged in discontinuous activities while sitting. 
\textit{Smartwatch-based solutions have shown promise with learning techniques, yet their reliability is compromised by motion artifacts, as on-the-wrist PPG and IMUs are highly susceptible to such disturbances~\cite{ModelingOfArtifactsInTheWristPhotoplethysmogram,electronics12132923,hao2017mindfulwatch}, leading to low data retention rates or high estimation errors, particularly under moving conditions. In contrast, \systemname introduces an earable-based training-free solution that achieves reliable RR monitoring in sedentary and active conditions without discarding any data.}

\textbf{Smartphone.} 
\cite{EstimationOfRespiratoryRatesUsingTheBuiltin, wang2021smartphone, nam2016monitoring, aly2016zephyr, rahman2020instantrr,SmartphoneMovementSensorsForTheRemoteMonitoring} employ sensors on smartphones for RR monitoring. \cite{rahman2020instantrr,aly2016zephyr,SmartphoneMovementSensorsForTheRemoteMonitoring} needs the user holds the smartphone or smartwatch against the chest~\cite{aly2016zephyr,rahman2020instantrr,SmartphoneMovementSensorsForTheRemoteMonitoring} or abdomen~\cite{rahman2020instantrr}, respectively, utilizing IMUs to capture breathing movements. \cite{EstimationOfRespiratoryRatesUsingTheBuiltin} records breathing sounds by placing the built-in microphones and headset microphones of the smartphone near the suprasternal notch and nose, respectively. \cite{nam2016monitoring} stabilizes the phone camera to face the user's chest for capturing breathing movements. \cite{wang2021smartphone} utilizes active acoustic sensing to monitor breathing-induced chest movements, requiring the smartphone to be held or placed in a specific posture. 
\textit{These solutions can only operate under stationary conditions and preclude continuous and real life monitoring due to the requirement of active user involvement.}

\textbf{Others.} 1) Wireless signals, such as RF~\cite{ExtractingMultiPersonRespirationFromEntangledRFSignals,SmartHomesThatMonitorBreathingAndHeart} and WiFi~\cite{WiPhoneSmartphonebasedRespirationMonitoringUsingAmbientReflected}, are utilized for RR monitoring, and have achieved significant success \textit{under stationary conditions, such as being still~\cite{WiPhoneSmartphonebasedRespirationMonitoringUsingAmbientReflected}, sleeping~\cite{ExtractingMultiPersonRespirationFromEntangledRFSignals}, and sitting~\cite{SmartHomesThatMonitorBreathingAndHeart}}. 2) Zephyr~\cite{Zephyr} employs pressure sensors in chest straps to capture breathing movements. \textit{However, such on-body devices can be intrusive and uncomfortable for users.} 3) Various types of nose-worn sensors~\cite{ContactBasedMethodsForMeasuringRespiratoryRate} are used to monitor the airflow through the nose and mouth during breathing. \textit{However, they are invasive and generally have low acceptance among users.} 4) RGB cameras and infrared thermography~\cite{ContactlessVitalSignsMeasurementSystemUsingRGBThermal} sense the temperature changes caused by breathing. \textit{However, these methods are effective only under stationary conditions and also raise privacy concerns.} 5) Studies employing body-worn PPG and ECG sensors \cite{natarajan2021measurement, charlton2017breathing, karlen2011respiratory, xu2022toward, schafer2008estimation} estimate RR indirectly from heart signals, \textit{which require the user to remain still.}

\subsection{Earable-based RR monitoring}

\textbf{Photoplethysmogram (PPG):} \cite{EarablePOCERDevelopmentOfPointofCareEarSensor, OptiBreatheAnEarablebasedPPGSystemForContinuous, DevelopmentOfVitalSignsMonitoringWirelessEara} employ in-ear PPG for RR monitoring, but are only functional under stationary conditions. Additionally, \cite{EarablePOCERDevelopmentOfPointofCareEarSensor} requires controlled breathing at specific rates for optimal operation. \cite{InEarPPGForVitalSigns} investigates the use of in-ear PPG for RR estimation in various user activities, \ie stationary, talking, walking, and running. However, the accuracy is significant affected by physical activity, with error rates up to approximately 31\% under motions, due to the lack of a specific design for motion-resilient estimation. Moreover, PPG is not commonly found in commercial earables, unlike microphones which are commonly integrated. 
\textbf{IMUs:} \cite{TowardsRespirationRateMonitoringUsingAnInEar, TowardsMotionAwarePassiveRestingRespiratoryRateMonitoring, RRMonitorResourceAwareEndtoEndSystemForContinuousMonitoring} use IMUs on earphones to estimate RR, but they are effective only under stationary conditions by discarding data from periods with motions.
\textbf{Microphones:} \cite{InEarAudioWearableMeasurementOfHeartAnda} utilizes in-ear microphones on earphones to determine RR but only works for high-intensity breathing when the user is stationary, as natural breathing is sometimes imperceptible. \cite{EstimatingRespiratoryRateFromBreathAudioObtaineda} estimates RR using out-ear microphones on AirPods, employing deep learning techniques. It solely relies on audible breathing sounds, retained through perceptual annotation for model training and testing, which means it only works effectively for heavy breathing. Moreover, out-ear microphones are inherently vulnerable to environmental noises as breathing sounds are weak and attenuate significantly in air.
\textbf{Multiple sensors:} \cite{hernandez2014bioglass} estimates RR using IMUs and/or a camera on a head-mounted device. However, it requires the user to remain still for accurate measurement. \cite{RemoteBreathingRateTrackingInStationaryPosition} employs out-ear microphones and IMUs on earphones for estimating RR, tailored for stationary conditions involving head motion. 
\cite{RunBuddySmartphoneSystemForRunningRhythmMonitoring,gu2017detecting} utilize out-ear microphones on earphones and IMUs on smartphones to estimate RR-related factor - LRC, when the user is running. This can, in turn, be used for RR estimation. However, these works assume that the LRC remains constant during one estimation window, which makes the system less robust in daily settings.
 

%% file: 08-conclusion.tex
\section{Conclusions}

This paper presented \systemname, the first earable system for continuous, non-invasive, and reliable RR monitoring across both sedentary and active conditions. 
\systemname employs in-ear microphones and leverages unique relationships of our cardiovascular, gait and respiratory systems to optimize RR detection. 
We implemented \systemname prototype and conducted extensive experiments to evaluate its performance. The results demonstrate that \systemname outperforms the state-of-art, and is robust in a variety of contexts.